\documentclass[a4paper,11pt]{article}
\usepackage{jheppub} 
\usepackage{lineno,braket}
\usepackage{braket,soul}
\usepackage{ulem}
\usepackage{amsmath,cleveref}
\usepackage{amsfonts, amsthm}
\usepackage{mathrsfs}
\usepackage{verbatim}
\usepackage{amssymb}
\usepackage{hyperref}
\usepackage{inputenc, array,subfig}
\usepackage{textcomp}
\usepackage{appendix}
\usepackage{epsfig}
\usepackage{graphicx}
\allowdisplaybreaks
\usepackage{color}
\newcommand{\e}{\epsilon}
\newcommand{\be}[1]{\begin{equation}\label{#1} }
\newcommand{\ee}{\end{equation}}
\newcommand{\bea}[1]{\begin{eqnarray}\label{#1} }
\newcommand{\eea}{\end{eqnarray}}
\newcommand{\p}{\partial}

\renewcommand{\d}{\partial}

\renewcommand{\t}{\tau}
\newcommand{\s}{\sigma}
\definecolor{darkblue}{rgb}{0.0, 0.2, 0.6}

\newcommand{\nn}{\nonumber}
\renewcommand{\p}{\mathfrak{p}}

\title{High energy scattering and null strings}

\author{Arjun Bagchi, Sachin Grover, Sharang Rajesh Iyer, Amartya Saha.}
\affiliation{Indian Institute of Technology Kanpur, Kalyanpur, Kanpur, Uttar Pradesh 208016, India}

\emailAdd{abagchi, sgrover, siyer, amartyas@iitk.ac.in}

\abstract{We propose an instrinsic worldsheet description of the ultra-high energy regime of string scattering based on worldsheet symmetries. At very high energies, the fundamental string becomes tensionless and in flat target spacetimes, the worldsheet becomes a null surface. Tensionless null strings thus emerge and the worldsheet symmetries morph from two copies of the Virasoro algebra to the two dimensional (2d) conformal Carroll or equivalently the 3d Bondi-van der Burgh-Metzner-Sachs (BMS) algebra. Tensionless strings have three inequivalent vacua over which they can be constructed, leading to distinct quantum theories. High energy tensile strings naturally connect to null strings built on the so-called induced vacuum. 

\medskip

Our principle goal in this paper is the construction of scattering amplitudes for null strings in the induced vacuum. We show that these amplitudes, constructed from worldsheet methods of the null string, coincide with the high energy limit of usual string amplitudes. A crucial part of our analysis is the construction of integrated vertex operators. This achieved by relying on lessons from the parent string theory and following the tensionless limit carefully. A striking feature of the null string is the blurring of differences between open and closed strings. We see this at the level of the amplitudes as well. We then focus on four-point amplitudes and recover all expected regimes including the Gross-Mende regime and the Regge limit. We finally comment on a new class of vertex operators which arise naturally only in the zero tension string. This reproduces all our earlier analyses when put onshell but also has hints of signatures beyond the perturbative tensile string.}

\begin{document}
\maketitle
\flushbottom

\section{Introduction}
Understanding the quantum nature of gravity remains theoretical physics' outstanding challenge. Of the competing frameworks for addressing this question, string theory has been one of the most best candidates. One of the most endearing features of string theory is that it has very few tunable parameters, one of which is the length of the fundamental string $\ell_s$. In terms of the tension $(T)$ of the string, we have: 
\begin{align}\label{tension}
    T = \frac{1}{2 \pi \ell_s^2} = \frac{1}{2 \pi \alpha'}.
\end{align}
On grounds of consistency, the limit where the fundamental string shrinks to a point should reproduce known physics from string theory and this $\alpha'\to0$ limit is well studied and is designed to give us supergravity from superstring theory. 
While this is the point-particle limit, the very stringy nature of string theory and hence the very quantum nature of quantum gravity should manifest itself in the diametrically opposite limit $\alpha'\to \infty$. This is the limit we are interested in investigating in our paper. 

\medskip

\subsection*{Strings at high temperatures and high energies}
 String theory has an exponential growth of states. The density of states at a given energy $\rho(E) \sim e^{\beta_H E}$, where $\beta_H= \mathcal{T}^{-1}_H$ is the inverse of the Hagedorn temperature. The partition function for string theory thus diverge beyond temperatures $\mathcal{T}=\mathcal{T}_H$. Beyond this very high temperature, the theory is conjectured to undergo a phase transition into a Hagedorn phase with fundamentally different degrees of freedom where long strings emerge \cite{Atick:1988si, Giddings:1989xe, BOWICK1989631}.  

\medskip

This very high temperature and high energy regime of string theory seems to hold a lot of promise for understanding quantum gravity. The framework of string field theory seems to be the better way of approaching the problem. But in a first quantized version of string theory, can we say anything about this sector? We seem to be in a symmetry broken phase and wanting to answer questions of the whole theory without having access to it. 

\medskip

The situation, as understood famously by Gross (and Mende) \cite{Gross:1987kza, Gross:1987ar, Gross:1988ue, Gross:1989}, is similar to spontaneously broken gauge theories where one can attempt to get information of the unbroken symmetry of the entire theory by looking at the high energy behaviour of scattering amplitudes of the effective low energy theory. Gross and Mende \cite{Gross:1987kza, Gross:1987ar} considered the very high energy limit  ($\alpha'\to \infty$)  of string scattering and found remarkable ultra-soft behaviour of these amplitudes that is very different from usual quantum field theory (QFT). More concretely, the fixed angle scattering in string theory is exponentially suppressed in stark contrast with the power-law fall-off of usual relativistic QFTs. In subsequent work, Gross \cite{Gross:1988ue} also found an infinite number of linear relations between these high energy string scattering amplitudes. All of these hint at a very rich structure of this high energy regime and hence of new undiscovered symmetries of string theory when considered in its unbroken phase. 

\subsection*{Our goal in this paper}
The analysis of Gross and Mende, as we will review in the next section of our paper, was primarily an $\alpha'\to \infty$ limit on the string scattering amplitudes and an equivalent limit on the path integral around the leading saddle. Although there have been numerous follow up works, an  {\textit{intrinsic worldsheet formulation}} of this extreme high energy limit has not been achieved in the literature. This is the very important question we address in our current work. We will employ recent techniques developed for tensionless null strings and Carrollian symmetries that emerge on their worldsheet for formulating the high energy strings and their scattering.

\subsection*{Tensionless null strings}
The $\alpha'\to\infty$ limit, as we have seen in \eqref{tension}, is the tensionless limit of string theory. This is similar in spirit to the massless limit of point particles. Massless point particles travel on null geodesics. Tensionless strings similarly sweep out null worldsheets in the target space and hence (in flat embedding spacetimes) tensionless strings are synonymous with null strings \cite{Schild:1976vq}. In what follows, we will be using these terms interchangably. 

\medskip

It is well understood that taking the mass to zero in the point-particle worldline action $S = -m \int d\tau$ does not yield the massless point particle action. One similarly has to work more to get to the tensionless action from the usual tensile action. The action for the null string we will focus on in this paper is the so-called ILST action \cite{Isberg:1993av}:
\begin{align}
    S_{ILST} = \frac{1}{2\pi c'} \int d^2\sigma V^a V^b \partial_a X^\mu \partial_b X^\nu \eta_{\mu\nu}.
\end{align}
Here $V^a$ are vector densities that replace the worldsheet metric. $\eta_{\mu\nu}$ is the background metric. $c'$ is a constant dimensionally equivalent to $\alpha'$. We will explain the procedure to get this briefly in our review of the tensionless string in the coming sections. For more details and a taste of the history of null strings, the reader is pointed to the recent review \cite{Bagchi:2026wcu}. 

\medskip

One of the most striking features behind the recent revival of the study of tensionless null strings is the identification \cite{Bagchi:2013bga, Bagchi:2015nca} of the worldsheet symmetries with the Carrollian conformal algebra or equivalently the Bondi-van der Burgh-Metzner-Sachs algebra \cite{Bondi:1962px, Sachs:1962wk, Barnich:2006av}. This algebra replaces two copies of the Virasoro algebra on the worldsheet as it becomes null in the tensionless limit. 

\medskip

Carrollian symmetries, first discovered as a curious speed of light $c\to0$ limit of Poincare symmetries \cite{Leblond65, SenGupta:1966qer}, has seen a recent explosion of interest with applications ranging from condensed matter physics \cite{Bidussi:2021nmp, Bagchi:2022eui,Ara:2024fbr,Biswas:2025dte}, ultra-relativistic hydrodynamics \cite{Bagchi:2023ysc} to black hole horizons \cite{Penna:2018gfx,Donnay:2019jiz} and near horizon regions \cite{Bagchi:2026qpi}, holography for asymptotically flat spacetimes (AFS) \cite{Bagchi:2010zz, Bagchi:2012xr, Bagchi:2014iea, Bagchi:2015wna, Bagchi:2016bcd, Bagchi:2022emh, Donnay:2022aba, Bagchi:2023fbj, Bagchi:2023cen, Saha:2023hsl, Saha:2023abr, Alday:2024yyj, Ruzziconi:2024kzo, Bagchi:2024gnn}. The reader is pointed to the reviews \cite{Bagchi:2025vri,Nguyen:2025zhg, Ruzziconi:2026bix} for a comprehensive overview of Carrollian physics and applications to AFS holography.  

\medskip

For the string, the tensionless limit is a worldsheet Carroll limit that contracts the Virasoro algebra to the BMS \cite{Bagchi:2013bga, Bagchi:2015nca}. This is achieved in the following way: 
\begin{align} \label{tless-lim}
    \frac{\alpha'}{c'} \to \frac{1}{\epsilon}, \quad \epsilon \to 0.  
\end{align}
This is equivalent to sending the worldsheet speed of light to zero and turning the tensile worldsheet to a null surface. Armed with the machinery used to understand Carrollian CFTs, the worldsheet organising principle of the tensionless string has become less mysterious allowing for investigations into the quantization of the theory. This has led to the discovery of many curious and non-intuitive features of the quantum null string \cite{Bagchi:2020fpr, Bagchi:2019cay, Bagchi:2020ats}, the foremost of which is the emergence of three distinct quantum theories for the bosonic null string starting with the ILST action \cite{Bagchi:2020fpr}. Of these three, we will be concerned mostly with the quantum theory built on the so-called \textit{Induced vacuum}. The nomenclature follows from the fact that the representations of importance for the underlying BMS algebra is the induced representation \cite{Barnich:2014kra, Campoleoni:2016vsh}. This is the one which will follow directly from the Carroll limit of the highest weight representation of the Virasoro algebra, which is the representation under which the tensile string states are organised. 

\subsection*{Tensionless strings and high energies}
So, why is the $\alpha'\to \infty$ limit a high energy limit? The essential point is to note what the dimensionless object can be constructed out of $\alpha'$. We see that  
\begin{align}
\alpha' E^2 = c' \mathfrak{E}^2
\end{align}
In the above, $E$ is an energy scale attached to the tensile string. This will be connected to the momenta of string states connected to the vertex operator one builds to understand string scattering. $\mathfrak{E}$ is an energy scale associated with an intrinsic null string. The tensionless limit, done in a proper dimensionless way is given by \eqref{tless-lim}. Hence 
\begin{align}
\boxed{\alpha' \to \frac{c'}{\epsilon}, \quad \epsilon \to 0 \, \Rightarrow \mathfrak{E} \to \frac{E}{\sqrt{\epsilon}}. }
\end{align}
So a tensionless limit is naturally a very high energy limit on the string as was to be expected from the fact that this is the opposite limit to the $\alpha'\to 0$ low energy point particle limit. We see later in the paper that the null string naturally comes with very high momenta associated with null string states, which again will be connected to the vertex operators we build in the paper.

\medskip

We have mentioned the exponential suppression of the high energy fixed angle scattering of string amplitudes. Gross and Mende attempted to give a geometric picture of this. They argued that in this case, the string becomes macroscopically stretched during the collision. The dominant configuration resembles a long folded string connecting the incoming and outgoing states. This picture explains the ultra-soft behaviour of string scattering.

\medskip

Interestingly, independently, it has been argued that in the tensionless limit, long folded open strings emerge from closed strings \cite{Bagchi:2019cay} and explicit expressions for boundary states emerging out of the worldsheet Carroll limit has been obtained \cite{Bagchi:2019cay, Bagchi:2020ats}. We will see further evidence of the Gross-Mende features naturally arising when we consider the scattering of null strings. 

\subsection*{Tensionless Vertex operators, Induced Vacuum and beyond}
One of our main accomplishments in this paper is the construction of vertex operators in the tensionless string. A class of vertex operators were previously defined in the highest weight vacuum (for the scalar Carrollian theory) in \cite{Hao:2021urq}. However, when constructing the intergrated form for these vertex operators to use for the null string in the highest weight vacuum, there were subtleties \cite{Chen:2025gaz}.  

\medskip

We will address a different question, that of finding vertex operators for the induced vacuum and we will see that we can find perfectly proper vertex operators which give integrated vertex operators and generate scattering amplitudes for the null string. 

\medskip

The induced vacuum often comes with limited information about operator ordering as we will learn in the following sections. We re-emphasise that we wish to calculate the very high energy sector of usual string theory. So we will be interested in the theory that emerges in the tensionless limit of usual tensile string theory. To this end, whenever the intrinsic description may obscure required information, we will fall back to usual string theory and define the required information as a limit of the parent tensile theory. 

\medskip

The reader may be apprehensive that by following this route, i.e. by limiting ourselves to only the analysis from the tensionless limit, we are ignoring some interesting physics that the purely null string may give us. At the end of the paper, we will introduce a class of vertex operators that seem to be intrinsic to the null string and only appear in the strict limit when the theory has gone completely null. While we will defer a detailed investigation of this vertex operator to future work, we reproduce our results earlier in the paper by putting this vertex operator ``on-shell''. Perhaps this class of vertex operators would become important when we attempt to explore beyond the Hagedorn temperature into the mysterious phase of strings. We will revisit some of these comments at the end of our paper. 

\subsection*{Outline of the paper}
The rest of the paper is structured as follows. In section \ref{sec: high-energy-strings}, we review the high energy scattering of open and closed strings. In section \ref{sec: review-nullstrings}, we review classical and quantum aspects of open and closed null strings. In the same section, we also discuss the quantisation of null-strings and key features of the induced vacuum. 

\medskip

Section \ref{BMS correlators from the Carroll limit} focuses on limiting analysis. We use the relation between the plane coordinates in 2d CFT ($z,\bar z$) and 2d CCFT ($t,x$) to take the Carroll limit on various 2d CFT correlators to obtain the corresponding correlators in 2d CCFT. We highlight several key features of the ultra-relativistic limit in this section. 

\medskip

In section \ref{sec: intrinsic-analysis}, we discuss the intrinsic Carroll worldsheet analysis for the free boson. This section is inspired by the limiting analysis and fine-tuned for the induced vacuum. Here we propose an ordering for the $A_n$-modes. This ordering respects the commutators of the modes and reproduces the results obtained from the limiting analysis. We also calculate the vertex operator correlators. 

\medskip

In section \ref{sec: scattering}, we define integrated vertex operators, and impose upon them diffeomorphism invariance to obtain the on-shell condition on the momenta. We compute the tree-level three and four-point tensionless string scattering amplitudes and comment on its relation to the Gross-Mende saddle. 

\medskip

Section \ref{sec: general-V_p,zeta} focuses on a new class of vertex operators. We calculate their correlators, define their integrated versions and comment on the scattering amplitude obtained from these new vertex operators. We conclude with some discussions and a list of future directions in section \ref{Conclusion and discussion}. 

\medskip

The appendices are organised as follows. In appendix \ref{app: measure}, we compute how the measure of the integrated vertex operators transforms under BMS transformations. In appendix \ref{app: AB-normal-order}, we explicitly compute the vertex operator correlators using the normal ordering prescription tailored to the induced vacuum. In appendix \ref{app: BMS-weights}, we prescribe a procedure to compute the BMS weights of the vertex operators in the induced vacuum. Finally, appendix \ref{app: closed-type-integrated-Vp} focuses on the scattering amplitude of closed tensionless strings and some related issues. 

\bigskip \bigskip

\section{High energy string scattering review}\label{sec: high-energy-strings}

The high energy scattering amplitudes in string theory exhibit remarkable universal features at every order in string perturbation theory. The high energy behaviour of the amplitude is independent of the string scattering states or the type of string theory \cite{Gross:1987ar, Gross:1987kza}. Importantly for us, the string scattering amplitudes are analytic functions of string tension or $\alpha'$, allowing a smooth, well-defined tensionless limit $\alpha'\to \infty$ of the scattering amplitudes. As we discussed, the tensionless limit focusses on the high energy sector of string theory. In this section, we review salient features of the high energy limit of string amplitudes, which we will later reproduce in the paper with an analysis based on the worldsheet of tensionless strings. 
\medskip

String amplitudes are dictated by the underlying worldsheet conformal symmetry. The simplest non-trivial amplitude is the tree-level 2-2 scattering amplitude of tachyons in bosonic string theory, which for open strings, takes the simple form,
\begin{multline}\label{eq: tensile-open-amplitude}
\mathcal{A}^4_{D_2}=\frac{2ig_0^2}{\alpha'}(2\pi)^{26}\delta^{26}\left(\sum_i^4 p_i\right)\left[\frac{\Gamma(-s-1)\Gamma(-t-1)}{\Gamma(-s-t-2)}\right.\\
+\left.\frac{\Gamma(-s-1)\Gamma(-u-1)}{\Gamma(-s-u-2)}+\frac{\Gamma(-u-1)\Gamma(-t-1)}{\Gamma(-u-t-2)}\right]\,.
\end{multline}
This is the \textit{Veneziano amplitude}. The kinematic Mandelstam variables are defined with $\alpha'$ to make them dimensionless,
\begin{equation}
    s=-\alpha'(p_1+p_2)^2\, ,\quad t=-\alpha'(p_1+p_3)^2\, ,\quad u=-\alpha'(p_1+p_4)^2\, .
\end{equation}
The Veneziano amplitude is manifestly crossing symmetric, is Regge bounded and consistent with unitarity among a host of interesting properties. The amplitude is invariant under diffeomorphisms of the worldsheet if the tachyons are on-shell,
\begin{equation}
   \alpha' p_i^2=1\, , \quad 1\leq i\leq 4\, .
\end{equation}
To extract the high energy behaviour of the amplitude we can use the Stirling's formula,
\begin{equation}\label{eq:Stirling's Formula}
        \Gamma(z)=(2\pi)^{1/2}e^{-z}z^{z-1/2}\exp\left(-\sum_{k=1}^\infty\frac{\zeta(1-2k)}{2k-1}z^{1-2k}\right)\,,
\end{equation}
for each of the Gamma function factor in the amplitude \eqref{eq: tensile-open-amplitude} together with the massless on-shell condition, $s+t+u\simeq 0$. There are still two ways to take the high energy limit on the 2-2 scattering amplitudes. In the centre of momentum frame, the two limits are the finite scattering angle or the infinitesimal scattering angle limits, also known as the \textit{Gross-Mende limit} and the \textit{Regge limit} respectively.
\begin{enumerate}
    \item Hard scattering limit $s\to\infty$, $u/s$ fixed or the \textit{Gross-Mende limit} leads to 
\begin{equation}\label{eq: open-Gross-Mende-limit}
\mathcal{A}^{(4)}(\{p_i\})\sim \exp\left(-(s\log s+t\log t+u\log u)\right)\, ,
\end{equation}
 
\item The \textit{Regge limit} $s\to\infty$, $u$ fixed, yields:
\begin{equation}
    \mathcal{A}^{(4)}_{D_2}(\{p_i\})\sim s^{u+1}\Gamma(-u-1)\, ,
\end{equation} 
\end{enumerate}

Interestingly the 2-2 closed string \textit{Virasoro-Shapiro} amplitude, 
\begin{equation}
    \mathcal{A}^4_{S_2}=\frac{8\pi ig_s^2}{\alpha'}(2\pi)^{26}\delta^{26}\left(\sum_i^4 p_i\right)\frac{\Gamma(-1-s/4)\Gamma(-1-t/4)\Gamma(-1-u/4)}{\Gamma(2+s/4)\Gamma(2+t/4)\Gamma(2+u/4)}\, ,
\end{equation}
with the on-shell condition, $s=4/\alpha'$ has similar high energy behaviour. The Gross-Mende limit is,
\begin{equation}
   \mathcal{A}^{(4)}(\{p_i\})\sim \exp\left(-\frac{1}{2}(s\log s+t\log t+u\log u)\right) \, .
\end{equation}
The factor of $1/2$ in the exponent, compared to the open string high energy amplitude \eqref{eq: open-Gross-Mende-limit} is most easily explained due to the \textit{doubling trick} \cite{Gross:1989}. Essentially, the open string amplitude is described by the vertex operator momenta that are doubled due to reflection about the horizontal axis of the plane while the area is halved.
\medskip

The Regge limit for the closed string tachyon 2-2 scattering is,
\begin{equation}
    \mathcal{A}^{(4)}_{S_2}(\{p_i\})\sim s^{2u+2}\frac{\Gamma(-1-u)}{\Gamma(2+u)}\, .
\end{equation}

The spectrum of high energy closed strings is again described by massless states, $p^2\simeq 0$ compared to the centre of mass energy of scattering.

We can convert the closed on-shell kinematic variables to the open on-shell variables. In the high energy limit we see the doubling relation between open and closed string tree-level amplitudes \cite{Kawai:1985xq}, 
\begin{equation} \label{eq: open-closed-square}
     \mathcal{A}^{(4)}_{S_2}\sim  \left(\mathcal{A}^{(4)}_{D_2}\right)^2\, .
\end{equation}

The explicit expression of the 2-2 scattering amplitude at tree level in string perturbation is well known. However, despite recent progress in one-loop string amplitudes for finite $\alpha'$, \cite{Eberhardt:2023xck, Baccianti:2025whd, Baccianti:2026lpc}, it is generally very hard to evaluate the integrals at any arbitrary order in perturbation theory. The high energy features of string amplitudes thus become vital and sometimes the only analytic result at an arbitrary order in string perturbation theory. The high energy behaviour of the string amplitudes at any order in perturbation theory exhibits the common dominant asymptotic behaviour described by the Gross-Mende saddle \cite{Gross:1987kza} of the bosonic string path integral. The $n$-point tachyon amplitude at an arbitrary perturbation order is given by the insertion of the integrated vertex operators $V_p$ in the bosonic path integral,
\begin{equation}
    \mathcal{A}_{\mathcal{M}}(\{p_i\})=\frac{g^{n-\chi}}{V}\int_\mathcal{M} \mathcal{D} g\, \mathcal{D}X \exp{\left(-\frac{1}{4\pi \alpha'}\int \sqrt{-\gamma} \gamma^{ab} \partial_aX^\mu \partial_bX^\nu \eta_{\mu\nu}\prod_{i=1}^{n}V_{p_i}\right)}\, ,
\end{equation}
where $V$ is the volume of the group of diffeomorphisms times Weyl transformations, $\chi$ is the Euler number of the topology of the genus $G$ Riemann surface $\mathcal{M}$. $V_{p_i}$ can either be the open or closed integrated vertex operators for the $i$-th on-shell tachyon with spacetime momentum $p_i$ respectively,
\begin{equation}
    V_{p_i}^o=\int_{\d \mathcal{M}} dx\sqrt{g} e^{ip\cdot X}\, ,\quad 
    V_{p_i}^c=\int_{\mathcal{M}} d^2x\sqrt{g} e^{ip\cdot X}\, .
\end{equation}

The 2-2 tachyon scattering amplitude is simply obtained by evaluating the Gaussian integral \cite{Gross:1987kza},\footnote{The exponent in the open string has a factor of 2 due to the doubling trick as discussed above.}
\begin{equation}\label{eq: def-amplitude-saddle}
    \mathcal{A}^{(4)}_G(\{p_i\})=\frac{g^{4-\chi}}{V}\int d\mathbf{m} \prod_{i=1}^{4}d^2x_i\,\Omega(m,x)\sqrt{g(x)}\exp\left(-\sum_{i<j}p_ip_j\braket{X(i)X(j)}_\mathcal{M}\right)\, ,
\end{equation}
where $\Omega$ is the standard measure of the moduli space $\mathcal{M}$ with moduli parameters $\mathbf{m}$ and $\braket{X(i)X(j)}_\mathcal{M}$ is the Green's function on $\mathcal{M}$. At tree-level, the above amplitude is the same as the correlator in the flat gauge,\footnote{The amplitude is only proportional since the integral is divergent due to the presence of poles. To compute the amplitude, three of the four positions are fixed, and we divide by the volume of the group of global diffeomorphisms.}
\begin{equation}\label{eq: open-amplitude-def1}
     \mathcal{A}^{(4)}_{D_2}\sim\frac{g_o^2}{\text{Vol} (SL(2,\mathbb{R}))}\int dz \braket{\mathcal{V}_{p_1}(0)\mathcal{V}_{p_2}(z)\mathcal{V}_{p_3}(1)\mathcal{V}_{p_4}(\infty)}\, .
\end{equation}
and for the closed string,
\begin{equation}\label{eq: closed-amplitude-def1}
    \mathcal{A}^{(4)}_{S_2}\sim\frac{g_s^2}{\text{Vol} (SL(2,\mathbb{C}))}\int dz d\bar z \braket{\mathcal{V}_{p_1}(0)\mathcal{V}_{p_2}(z)\mathcal{V}_{p_3}(1)\mathcal{V}_{p_4}(\infty)}\, .
\end{equation}
where $\bar z$, $ z$ are the (anti-)holomorphic cross-ratios. 
\medskip

The two-point function on the sphere of the scalars is,
\begin{equation}
    \braket{X^\mu(z_1,\bar{z}_1)X^\nu(z_2,\bar{z}_2)}_{S^2}=-\frac{\alpha'}{2}\log((z_1-z_2)(\bar{z}_1-\bar{z}_2))\eta^{\mu\nu}\, .
\end{equation}
The tree-level closed string amplitude is thus given by the integral over the cross-ratio,
\begin{equation}
    \mathcal{A}^{(4)}_{S^2}\sim g_s^2\int d^2 z \, \exp\left(-\frac{1}{2}\left((t+16)\log|z|+((u+16)\log|1-z|\right)\right)\, .
\end{equation}
The Gross-Mende saddle is described by the limit in which the states are massless $p_i^2\sim 0$ compared to the product, $p_i\cdot p_j$,
\begin{equation}
    \mathcal{A}^{(4)}\sim g_s^2\int d^2 z \, \exp\left(-\frac{1}{2}\left(t\log|z|+(u\log|1-z|\right)\right)\, .
\end{equation}
For the above amplitude the saddle occurs in the middle of the modulus space at,
\begin{equation}
    z_{s}=-\frac{t}{s}\, ,\quad 1-z_s=-\frac{u}{s}\, .
\end{equation}
The tree-level amplitude is thus dominated by,
\begin{equation}
   \mathcal{A}^{(4)}(\{p_i\})\sim \exp\left(-\frac{1}{r}(s\log s+t\log t+u\log u)\right) \, ,
\end{equation}
where $r=1$ for the open and $r=2$ for the closed string with open and closed Mandelstam variables respectively.
\medskip

Recently, there has been a revival of interest in the high energy behaviour of string amplitudes due to its connection with celestial holography \cite{Stieberger:2018edy, Kervyn:2025wsb}. In this regard, the Gross-Mende behaviour of the Veneziano and the Virasoro-Shapiro amplitudes was derived a bit more carefully in \cite{Stieberger:2018edy} and revisited in \cite{Kervyn:2025wsb} where the authors realised the need to analytically continue the Gamma-functions to the physical region of the Mandelstam variables, $\Re(s)\to \infty$ and $\Re(t)\, ,\Re(u)\to -\infty$. Although the pole structure and the asymptotic form of the amplitude remain the same, the open string amplitude has a phase factor $(-1)^{-u-t}$ when the Gamma-functions are expanded in the physical region compared to the naive factor $(-1)^{-s-u}$ expected from \cite{Gross:1987kza}. 
\medskip

\section{Null strings: Closed and open}\label{sec: review-nullstrings}

We provide a brief review of bosonic tensionless strings in this section. For a more detailed account, we refer the reader to the recent review \cite{Bagchi:2026wcu}. The tensionless string theory action was derived in \cite{Isberg:1993av} by taking the tensionless limit of bosonic string theory. The tensionless string has a null worldsheet where the residual symmetries combine to form an infinite- dimensional BMS$_3$ residual symmetry algebra which is isomorphic to the CCFT$_2$ algebra \cite{Bagchi:2010zz, Duval:2014uva, Bagchi:2013bga}. We treat the tensionless string action as the starting point of the intrinsic analysis of null strings. 

\subsection{Classical null string}
The  tensile bosonic string is governed by the Polyakov action, 
\begin{align}\label{Polyakov action}
    S_{P} = - \frac{T}{2} \int d^2\xi \sqrt{-\gamma} \gamma^{ab} \partial_aX^\mu \partial_bX^\nu \eta_{\mu\nu}.  
\end{align}
Here $\gamma^{ab}$ is the worldsheet metric and the string propagates in a flat $D$-dimensional target spacetime with the metric $\eta_{\mu\nu}$. The string tension is denoted by $T$. A naive $T\to0$ limit renders the Polyakov action trivial. The null strings are the extended analogues of massless point particles. The massless limit of point particles extends to the tensionless limit in string theory. Classically, the fundamental string becomes long and floppy in the tensionless limit $T\to 0$. This manifests itself on the worldsheet by making the 2d surface null \cite{Bagchi:2013bga, Bagchi:2015nca}. In terms of an action, the null tensionless string action is given by the action \cite{Isberg:1993av},
\begin{align}\label{ILST-action}
    S_{ILST} = \int d^2\xi \, V^a V^b \partial_aX^\mu \partial_bX^\nu \eta_{\mu\nu}. 
\end{align}
where the following effective replacement has occurred in the tensionless limit, 
\begin{align}
   -\frac{T}{2} \sqrt{-\gamma} \gamma^{ab} \xrightarrow{T\to0}  V^a V^b. 
\end{align}
The worldsheet metric $\gamma^{ab}$ becomes a rank one null metric in the tensionless limit and is replaced by the vector density $V^a$. The above action can also be reached through a Hamiltonian formulation \cite{Isberg:1993av, Bagchi:2026wcu}. 

\paragraph{Symmetries:} The ILST action is invariant under worldsheet diffeomorphisms,
\begin{equation}\label{eq: BMS3-diff}
    \xi^{a}\to\xi^a+\epsilon^a\, ,
\end{equation}
where the scalar field and vector density transform as,
\begin{equation}
    \delta X^\mu=\epsilon\cdot\partial X (\xi)\, , \quad  \delta V^a=V\cdot\partial\epsilon^a-\epsilon\cdot\partial V^a+\frac{1}{2}\partial\cdot\epsilon V^a\, .
\end{equation}
The conformal gauge $\gamma^{ab}=e^\phi \eta^{ab}$ in the Polyakov action corresponds to the transverse gauge on the cylinder worldsheet,
\begin{equation}
    V^a=(v,0)\, ,
\end{equation}
where $v$ is a constant. We set $v=1$ following \cite{Bagchi:2020fpr}. The residual worldsheet symmetry which leaves the gauge-fixed action invariant is,
\begin{equation}
   (\tau',\sigma')=(f'(\sigma)\tau +g(\sigma), f(\sigma))\, ,
\end{equation}
where $f(\sigma)$ and $g(\sigma)$ are independent functions of the spatial coordinate $\sigma$ and $\tau$ is the non-compact coordinate on the cylinder. The generators of the symmetry act on an arbitrary function $F(\xi)$ of the coordinates as,
\begin{align}
L(F)&=\left(f'(\sigma)\tau\partial_\tau+f(\sigma)\partial_\sigma\right)F(\xi), \quad M(F)=(g(\sigma)\partial_\tau)F(\xi)\, .
\end{align}

The functions $f(\sigma)=\sum a_n e^{in\sigma}$ and $g(\sigma)=\sum b_n e^{in\sigma}$ can be expanded in Fourier modes to obtain the modes of the generators,
\begin{subequations}
    \begin{align}
L(\sigma,\tau)&=\sum_{n\in\mathbb{Z}} a_n e^{in\sigma}\left(\partial_\sigma+in\tau\partial_\tau\right)=-i\sum_{n\in\mathbb{Z}}a_nL_n\, ,\\
M(\sigma)&=\sum_{n\in\mathbb{Z}} b_ne^{in\sigma}\partial_\tau=-i\sum_{n\in\mathbb{Z}} b_n M_n\, .
\end{align}
\end{subequations}
The modes satisfy the classical BMS$_3$ algebra,
\begin{equation}\label{bms3}
    [L_n,L_m]=(n-m)L_{n+m}\,,\quad
    [L_n,M_m]=(n-m)M_{n+m}\,,\quad
    [M_n,M_m]=0\,.
\end{equation}
We will review how these residual symmetries are the key to understanding the tensionless string below. 

\paragraph{Equations of motion:} The equations of motion of the fields are,
\begin{equation}
    V^b(\partial_a X^\mu\partial_b X^\nu \eta_{\mu\nu})=0\, ,\quad \partial_a(V^aV^b\partial_b X^\mu)=0\, .
\end{equation}
The metric on the worldsheet $\gamma_{ab}=\partial_a X\cdot  \partial_b X$ is thus degenerate on-shell. In the transverse gauge, the equations of motions reduces to,
\begin{equation}\label{eq: eom-X}
    \partial^2_\tau X^\mu=0\, ,\quad 
    (\partial_\tau X)^2=0\, ,\quad \partial_\tau X\cdot \partial_\sigma X=0\, .
\end{equation}
The first of the above equations describes the motion of the classical string, while the remaining two equations act as constraints on the classical motion of the string. Physically, the classical null string in the transverse gauge has a degree of freedom in the transverse direction to the string. We can then interpret the classical null string as a collection of massless point particles \cite{Isberg:1993av}, although it is very important to keep in mind that the constraints ``string'' together these massless point particles.  

\subsection{Mode expansion of the tensionless string}

We now discuss the quantum nature of the null string. The solution of the equation of motion for $X^\mu$ is,
\begin{equation}\label{eq: X-mode-expansion}
    X^\mu=x^\mu+\sqrt{\frac{c'}{2}}A_0^\mu\sigma+\sqrt{\frac{c'}{2}}B_0\tau+i\sqrt{\frac{c'}{2}}\sum_{n\neq0}\frac{1}{n}\left(A_n-in\tau B_n\right)e^{-in\sigma},\; 
\end{equation}
The closed string mode expansion is periodic in $\sigma$,
\begin{equation}
    X^\mu(\tau,\sigma)=X^\mu(\tau,\sigma+2\pi)\, ,
\end{equation}
which implies $A_0^\mu=0$ for closed strings. The open string has a boundary $\partial M$ on $\sigma=0$ and $\sigma=\pi$. The boundary term from varying the ILST action \eqref{ILST-action} is,
\begin{equation}
    \int_{\partial M} d\, n_{a}V^a V^b (\partial_b X^\nu)\delta X^\mu \eta_{\mu\nu}=0\, ,
\end{equation}
where $n_a$ is the normal to the boundary of the string $\partial M$. Apart from the usual Dirichlet ($\delta X^\mu|_{\sigma=0,\pi}=0$) and Neumann ($V^b\partial_b X^\nu|_{\sigma=0,\pi}=0$) boundary conditions,
we also obtain the \textit{null boundary condition} which is special to the null string:
\begin{equation}\label{nullbc}
    n_{a}V^a=0\, .
\end{equation}
The null boundary condition does not put any constraint on the mode expansion of the scalar field $X$ \eqref{eq: X-mode-expansion}, i.e. imposing the null boundary condition, $X$ has the mode expansion $\eqref{eq: X-mode-expansion}$ without any further constraints \cite{Bagchi:2019cay, Bagchi:2024qsb}. This is the boundary condition for the open string we would be interested in. We note that this is a very important feature of null strings. Open null strings subject to null boundary conditions are identical in terms of the ILST action to the closed string. We see that, giving credence to the folklore about tensionless strings, the distinctions between open and closed strings blur in the tensionless limit. 

\medskip

It is important to note here that taking the tensionless limit on the Dirichlet or Neumann string will lead to something fundamentally different from the null boundary condition. The residual symmetry algebra changes in these cases and is given by what is called the Boundary Carrollian conformal algebra, which can be systematically obtained either from the CCA \eqref{bms3} by imposing a relation between the modes, or by a novel contraction of a {\it{single copy of the Virasoro algebra}}. For details, the reader is referred to \cite{Bagchi:2024qsb,Bagchi:2025jgu}. We will not be interested in these open null strings and hence will keep working with just the ILST action.

\medskip

The Virasoro constraints now take the form $(\partial_\tau X)^2=0$, and $\partial_\tau X\cdot \partial_\sigma X=0$ on the tensionless strings,
\begin{equation}
    (\partial_\tau X)^2=\sum_n M_ne^{-in\sigma}=0\,,\quad
    \partial_\tau X\cdot \partial_\sigma X=\sum_n(L_n-in\tau M_n)e^{-in\sigma}=0
\end{equation}
results in the \textit{sandwich conditions} \cite{Bagchi:2020fpr} on the physical Hilbert space,
\begin{equation}
    \braket{\psi |L_n|\psi'}=0\, ,\quad \braket{\psi| M_n|\psi'}=0\, ,
\end{equation}
where states $\psi$ and $\psi'$ denote the set of all states in the physical Hilbert space.
The constraints can be satisfied by three different types of vacua \cite{Bagchi:2020fpr}. Here we have  \cite{Bagchi:2021rfw},
\begin{equation}\label{eq: L-M-mode-expansion}
    L_n=\frac{1}{2}\sum_{m}:A_{-m}\cdot B_{m+n}:\, \quad
    M_n=\frac{1}{2}\sum_{m}:B_{-m}B_{m+n}:\, .
\end{equation}

The $L_n$ and $M_n$ are the conserved charges corresponding to the residual symmetry of the ILST action \eqref{eq: BMS3-diff}. In terms of the components of the stress tensor, the charges, $L_n$ and $M_n$, are given as,
\begin{equation}\label{eq:charges-stress tensor}
    L_n=\int d\sigma(T^0_0 in\tau+T^0_1)e^{in\sigma}\, ,\quad M_n=\int d\sigma T^0_0 e^{in\sigma}\,.
\end{equation}
To obtain \eqref{eq:charges-stress tensor}, we have used the mode expansion \eqref{eq: X-mode-expansion}. 

\medskip

Imposing equal time canonical commutation relations, we get,
 \begin{equation}\label{eq: A-B-commutator}
[A^\mu_m,A^\nu_n]=0=[B^\mu_m, B^\nu_n]\, ,\quad [A^\mu_m,B^\nu_n]=2m\delta_{m+n}\eta^{\mu\nu}\,,\quad [x^\mu,B^\nu_0]=2i\sqrt{\frac{c'}{2}}\eta^{\mu\nu}\, , 
 \end{equation}
Here $B_0$ is the centre of mass momentum canonically conjugate to the position $x^\mu$ of the centre of mass. Using the above commutation relations, we find that the Noether charges satisfy the centrally extended BMS$_3$ algebra,
\begin{subequations}\label{eq: BMS3-algebra}
    \begin{align}
        [L_n,L_m]&=(n-m)L_{n+m}+\frac{c_L}{12}n(n^2-1)\delta_{n+m}\, ,\\
    [L_n,M_m]&=(n-m)M_{n+m}+\frac{c_M}{12}n(n^2-1)\delta_{n+m}\, ,\\
    [M_n,M_m]&=0\, .
    \end{align}
    \end{subequations}
Here $c_L, c_M$ are central charges which depend crucially on the choice of vacuum, as we will see below. Note that in the above, when considering the open null string with null boundary conditions, the mode $A_0$ is non-zero. However, demanding that the underlying symmetry algebra is BMS$_3$ signifies that $A_0$ commutes with every other mode. 

\subsection{Tensionless limit as a worldsheet Carroll limit} 
The tensionless limit of string theory is described by taking the string tension,
\begin{equation}
    T=\frac{1}{2\pi\alpha'}\to0\, ,
\end{equation}
which implies $\alpha'\to\infty$. Since $\alpha'$ is a dimension-full parameter, we define, 
\begin{equation}\label{eq: tensionless-alpha-prime}
    \alpha'=\frac{c'}{\epsilon}\, ,
\end{equation}
and take the limit $\epsilon\to0$ where $c'$ is a parameter which carries the dimensions of $\alpha'$. In the tensionless limit, the length of the string becomes infinite. Naively, apart from the boundary conditions, there is no difference between the classical open and closed string. The trick to impose closed string boundary conditions on the string of finite length is to consider the ultra-relativistic coordinate transformation,
\begin{equation}\label{eq: UR-limit-cylinder}
    \sigma\to\sigma\, ,\quad \tau\to\epsilon\tau\, ,
\end{equation}
where the tensionless regime is described in the limit $\epsilon\to0$. The interpretation of the tensionless string as an ultra-relativistic string becomes natural from this point of view. 

The string worldsheet action in the conformal gauge reduces to the free scalar action. We now wish to derive the Carroll equivalent of the free scalar action from a UR limit on the cylinder. We begin with the free scalar field action,
\begin{equation}
    S=\frac{1}{4\pi\alpha'}\int d\tau\, d\sigma~ \partial_\mu X \partial^\mu X\, ,
\end{equation}
and rescale the worldsheet coordinates, $\tau\to \epsilon \tau$, and the string tension $\alpha'= c'/\epsilon$, to get the leading piece of the action in the limit $\epsilon\to0$\, 
\begin{equation}\label{scalar-field-action-gen}
   S=\frac{1}{4\pi c'}\int d\tau\, d\sigma~(\partial_\tau X)^2.
   \end{equation}
The equation of motion is given by,
\begin{equation}\label{scalarfieldeom}
    \partial^2_\tau X(\tau,\sigma)=0\, .
\end{equation} 
We define a linear combination of the stress tensor, 
\begin{align}
    T_\sigma&=T^0_{1}-\tau\partial_\sigma T^{0}_{\, 0}\,, = \, :\partial_\tau X \partial_\sigma X:-\tau :\partial_\sigma\partial_\tau X\partial_\tau X:\, ,
\end{align}
which is also conserved, i.e., $\partial_\tau T_\sigma=0$ such that $T_\sigma= \sum_n L_n e^{-in\sigma}$. We rename the remaining component $T^0_0=T_\tau=\sum_n M_n e^{-in\sigma}$. The residual symmetry of the gauge fixed tensionless action is generated by an ultra-relativistic contraction of the Virasoro generators,
\begin{equation}\label{eq: BMS-in-terms-Vir-gens}
    L_n=\mathcal{L}_n-\bar{\mathcal{L}}_{-n}\, ,\quad 
    M_n=\epsilon\left(\mathcal{L}_n+\bar{\mathcal{L}}_{-n}\right)\, ,
\end{equation}
in the limit $\epsilon\to 0$, where $\mathcal{L}_n$ and $\bar{\mathcal{L}}_n$ generate the holomorphic and anti-holomorphic  Virasoro algebra. The $L_n$ and $M_n$ generate BMS$_3$ algebra given in \eqref{eq: BMS3-algebra} with,
\begin{equation}
    c_L=c-\bar c\, , \quad c_M=\epsilon(c+\bar c)\, .
\end{equation}
When $\epsilon\to0$, $c_M$ vanishes when the central charges of the tensile string theory are finite.
\medskip 

The match between tensionless limit of string theory and the intrinsic analysis serves as an important check for us. For example, the match between symmetries can be expanded to the mode expansion of the worldsheet boson. The tensile string theory free boson has the general mode expansion,
\begin{equation}\label{eq: X-alpha-mode-expansion}
    X_{\e=1}^\mu=x^\mu+\sqrt{\frac{\alpha'}{2}}\tau(\alpha_0^\mu+\tilde{\alpha}_0^\mu)+\sqrt{\frac{\alpha'}{2}}\sigma(\alpha_0^\mu-\tilde{\alpha}_0^\mu)+i\sqrt{\frac{\alpha'}{2}}\sum_{n\neq0}\frac{1}{n}\Big[\alpha_n^\mu e^{-in(\tau+\sigma)}+\tilde{\alpha}_n^\mu e^{-in(\tau-\sigma)}\Big]\, . 
\end{equation}
We can perform the UR limit \eqref{eq: tensionless-alpha-prime}, \eqref{eq: UR-limit-cylinder} in above to obtain \eqref{eq: X-mode-expansion} from the limiting analysis, when we identify,
\begin{equation}\label{eq: AB-to-alphas}
    A_n^\mu=\frac{1}{\sqrt{\epsilon}}(\alpha_n^\mu-\tilde{\alpha}^\mu_{-n})\, ,\quad B_n^\mu=\sqrt\epsilon(\alpha_n^\mu+\tilde{\alpha}^\mu_{-n})\, .
\end{equation}
The $A$ and $B$ modes satisfy the same algebra obtained by imposing canonical commutation relations \eqref{eq: A-B-commutator} or by using the commutators of the $\alpha$ oscillators,
\begin{equation}
[\alpha^\mu_n,\alpha^\nu_m]=n\delta_{n+m}\eta^{\mu\nu}\, ,\quad [\tilde \alpha^\mu_n,\tilde \alpha^\nu_m]=n\delta_{n+m}\eta^{\mu\nu}\,,\quad[x^\mu,\alpha_0^\nu]=[x^\mu,\tilde\alpha_0^\nu]=i\sqrt{\frac{\alpha'}{2}}\eta^{\mu\nu}
\end{equation}
and substituting in the commutator $[A_n,B_m]$. The tensionless mode expansion is thus equivalent to the leading behaviour of the free relativistic boson in the limit $\lim\epsilon\to 0$.

\subsection{The induced representation}\label{sec:The induced representation}

The BMS$_3$ algebra is a semi-direct sum of the holomorphic Virasoro algebra generated by $L_n$'s and an Abelian ideal generated by $M_n$'s. The representation induced from the ideal is the induced representation. The induced representation for the BMS group was constructed in \cite{Barnich:2014kra, Campoleoni:2016vsh}. 
\medskip

A special case of the induced representation was constructed for the bosonic tensionless strings by imposing quantum constraints on the physical Hilbert space in \cite{Bagchi:2020fpr}. The induced vacuum is, of course, invariant under the global part of BMS, and the rest of the physical states are defined by,
\begin{equation}
    L_n\ket{\psi}\neq 0\, ,\quad M_n\ket{\psi}=0\, ,\quad \forall n\neq 0\, .
\end{equation}
The highest weight vacuum of the bosonic tensile string theory goes over to the induced vacuum of the bosonic tensionless string theory \cite{Bagchi:2019cay, Bagchi:2020fpr}. To see this, remember that in the Virasoro algebra, a highest weight state is defined as
\begin{equation}
    \mathcal{L}_n\ket{h,\bar h}=0\, ,\quad \bar{\mathcal{L}}_n\ket{h,\bar h}=0\, ,\quad  \forall n>0\, .
\end{equation}
In the UR limit \eqref{eq: BMS-in-terms-Vir-gens}, we obtain, 
\begin{equation}
    \left(L_n+\frac{1}{\epsilon}M_n\right)\ket{h,\bar h}=0\, ,\quad
    \left(L_n-\frac{1}{\epsilon}M_n\right)\ket{h,\bar h}=0\, , \quad n>0
\end{equation}
In the limit $\epsilon\to0$ assume that the highest weight state maps to some state,
\begin{equation}
    \lim_{\epsilon\to0}\ket{h,\bar h}=\ket{\Delta,\xi}\,,
\end{equation}
then we get the condition,
\begin{equation}\label{eq: induced-rep-def-part-1}
    M_n\ket{\Delta,\xi}=0\, ,\quad \forall n\neq 0\, ,
\end{equation}
along with the action,
\begin{equation}
    M_0\ket{\Delta,\xi}=\xi\ket{\Delta,\xi}\, ,\quad L_0\ket{\Delta,\xi}=\Delta\ket{\Delta,\xi}\, .
\end{equation}
 The conditions  on the state $\ket{M,s}$ define the states in the induced representation. Thus, we see that the induced representation is the UR limit of the highest weight representation of the Vir$\oplus\overline{\text{Vir}}$ algebra. In particular, the highest weight tensile vacuum, which we will refer to as the $\alpha$-vacuum, goes over to the induced vacuum of the tensionless string. The induced vacuum of tensionless string theory is invariant under the global symmetry generators, which sets the weight,
 \begin{equation}
     M_0\ket{0}_I=0\, .
 \end{equation}
 Using \eqref{eq: induced-rep-def-part-1} we find \cite{Bagchi:2019cay},
\begin{equation}\label{eq: induced-vac-conditions}
    B_n^\mu\ket{0}_I=0\quad\forall n\in\mathbb{Z}\, ,
\end{equation} 
since the BMS generators in terms of the modes can be written as 
\begin{equation}\label{eq:BMS3 gen in modes}
    L_n=\frac{1}{2}\sum_{m\in\mathbb{Z}}:A_{-m}B_{m+n}:\,, \quad 
    M_n=\frac{1}{2}\sum_{m\in\mathbb{Z}}:B_{-m}B_{m+n}:\, ,
\end{equation}
where the normal ordering of the generators is defined as follows
\begin{equation}\label{eq: A-B-normal-ordering}
    :A_nB_m:=A_nB_m\, ,\quad \forall~ n,m\in\mathbb{Z}\, .
\end{equation}
We thus have the induced representation vacuum $\ket{0}_I$ as the global symmetry preserving vacuum defined by \eqref{eq: induced-vac-conditions}.

\medskip

On the other hand, the $\alpha$-vacuum of tensile string theory is defined by
\begin{equation}\label{eq: hw-vacuum}
    \alpha_n\ket{0}_\alpha=\tilde\alpha_{n}\ket{0}_\alpha=0\, ,\quad \forall~ n\in\mathbb{Z}_{\geq 0}\, .
\end{equation}
Using \eqref{eq: AB-to-alphas}, in terms of the $A_n$ and $B_n$ modes, the condition on the highest weight vacuum becomes,
\begin{equation}
    \left(\sqrt{\epsilon}A_n+\frac{B_n}{\sqrt{\epsilon}}\right)\ket{0}_{\alpha}=0\, ,\quad
    \left(-\sqrt{\epsilon}A_{-n}+\frac{B_{-n}}{\sqrt{\epsilon}}\right)\ket{0}_{\alpha}=0\, ,\quad n\in\mathbb{Z}_{\geq 0}\, .
\end{equation}
In the $\epsilon\to0$ limit we get the induced vacuum conditions \eqref{eq: induced-vac-conditions}. Thus, we see that the induced vacuum is the natural UR limit of the highest weight vacuum \cite{Bagchi:2019cay},
\begin{equation}
    \ket{0}_\alpha\mapsto\ket{0}_I\, .
\end{equation}
In fact, in \cite{Bagchi:2019cay} it was shown that the entire module over the highest weight vacuum in the tensionless string theory condenses to the induced vacuum in the high energy limit $\epsilon\to0$. In particular, states like
\begin{equation}
   \lim_{\epsilon\to0} \varepsilon_{\mu\nu}\alpha^{\mu}_{-n_1}\alpha^{\nu}_{-n_2}\ket{0}_\alpha\propto \ket{0}_I\,.
\end{equation}

The central charges in the induced representation were computed in \cite{Bagchi:2021rfw} as $c_L=c_M=0$ by sandwiching the commutators of the symmetry algebra modes. In \cite{Bagchi:2021rfw} it was found that the induced representation string does not have a critical dimension. In other words, the null string built on the induced representation can live in any dimension, and in particular is consistent in $D=26$, i.e. as a limit of usual bosonic tensile string theory.

\subsection{Tensionless string theory on the plane}

Although the worldsheet theory of the null string is naturally defined on the cylinder, an equivalent description can be obtained on the plane using the map \cite{Bagchi:2013qva, Hao:2021urq},
\begin{equation}\label{eq: cylinder-to-plane-coords}
    x=e^{i\sigma}\, ,\quad t=i\tau e^{i\sigma}\, .
\end{equation}
Under this coordinate transformation, the action is invariant,
\begin{equation}
    S=\frac{1}{4\pi c'}\int d\sigma d\tau (\d_\tau X)^2=\frac{1}{4\pi c'}\int d x dt (\d_t X)^2\, ,
\end{equation}
where the measure transforms as,
\begin{equation}
   -x^{2} d\sigma d\tau= \,dx dt\, .
\end{equation}

Using the cylinder to plane map, the UR limit on the complex plane coordinates is, 
\begin{equation}\label{eq: UR-limit-on-plane}
    z=e^{i\sigma+\epsilon\tau}=x\left(1+\epsilon\frac{t}{x}\right)+\mathcal{O}(\epsilon)\, ,\quad
    \bar z=e^{-i\sigma+\epsilon\tau}=x^{-1}\left(1+\epsilon\frac{t}{x}\right)\, .
\end{equation}

On the plane, the coordinate $x=e^{i\sigma}$ is a complex coordinate which takes values on a circle and $t$ is like a Lorentzian time coordinate. A radial ordering prescription can also be defined on the plane coordinate by an $\textit{x-ordering}$ defined in \cite{Hao:2021urq}. On the plane, the BMS$_3$ generators take the form
\begin{equation}
    l_n=x^{n+1}\d_x+(n+1)tx^n\d_t\, ,\quad m_n=x^{n+1}\d_t\, .
\end{equation}
It is the plane coordinates which are eigenfunctions of the generators and not the cylinder coordinates. Since $X(\sigma,\tau)=X(x,t)$ by virtue of it being a scalar, we can simply use the coordinate transformation \eqref{eq: cylinder-to-plane-coords}, and obtain,
\begin{equation}
    X(x,t)=x_0-i\sqrt{\frac{c'}{2}} B_0\frac{t}{x}+i\sqrt{\frac{c'}{2}}\sum_{n\neq0}\frac{1}{n}\left(\frac{A_n}{x^n}-nt \frac{B_n}{x^{n+1}}\right)\,. 
\end{equation}
The transformation of the scalar field $X(x,t)$ under $L_n$ and $M_n$ is
\begin{equation}\label{eq: X-BMS-transform}
    [L_n,X]=x^{n+1}\partial_xX+(n+1)x^nt\partial_tX\,,\quad[M_n,X]=x^{n+1}\partial_tX\,.
\end{equation}
Although the commutators look like that of a scalar primary field of weights $\Delta=\xi=0$, remember that the field $X$ is non-local.

\section{Correlators from Carroll limit and vertex operators}\label{BMS correlators from the Carroll limit}

The relativistic free scalar in 2d has conserved currents,
\begin{equation}
    J(z)=\d_z X(z,\bar z)\, ,\quad J(\bar z)=\d_{\bar z} X(z,\bar z)\, ,
\end{equation}
which are Virasoro primaries with conformal weights $(h,\bar h)=(1,0)$ and $(0,1)$, respectively. Apart from the conserved currents there is the family of infinite number of Virasoro primary vertex operators,
\begin{equation}
    \mathcal{V}_p(z,\bar z)=:\exp(ip X(z,\bar z)):\, ,
\end{equation}
with  conformal weights $h=\bar h=p^2\alpha'/4$, where the momentum $p\in\mathbb{R}$ for a unitary CFT$_2$. 
\medskip

Similar to the relativistic free boson CFT as the worldsheet of the tensile bosonic string, we have the free boson conformal Carroll field theory as the worldsheet of the bosonic tensionless string. The basic field is the worldsheet scalar $X$. We will construct currents and vertex operators from an ultra-relativistic limit of CFT$_2$ and compute their correlation functions in this section.

\subsection{BMS scalar correlators}

The two-point function of the relativistic free boson on the plane is
\begin{equation}
    \braket{X^\mu(z_1,\bar{z}_1)X^\nu(z_2,\bar{z}_2)}=-\frac{\alpha'}{2}\log((z_1-z_2)(\bar{z}_1-\bar{z}_2))\eta^{\mu\nu}\, .
\end{equation}

The high energy limit of the above correlator is the leading piece upon inserting the UR limit \eqref{eq: UR-limit-on-plane} along with $\alpha'=c'/\epsilon$ in the limit $\e\to0$. The leading behaviour of the two-point function is
\begin{equation}\label{eq: XX-prop-limiting}
    \braket{X^\mu(1) X^\nu(2)}=-\frac{c'}{2\epsilon}\log\left(2-\frac{x_1}{x_2}-\frac{x_2}{x_1}\right)\eta^{\mu\nu}\, .
\end{equation}
The $\braket{XX}$ two-point function on the plane matches with the analysis of \cite{Hao:2021urq}. Although the propagator looks divergent, it will play a crucial role in the subsequent sections.

The two-point function can be put in the form,
\begin{equation}\label{eq: 2pt-XX-plane}
    \braket{X^\mu(1) X^\nu(2)}=-\frac{c'}{\epsilon}\log(x_{12})+\frac{c'}{2\epsilon}\log(x_1 x_2)-i\pi\frac{c'}{\epsilon}\, .
\end{equation}
Notice that the second term spoils the translational symmetry along the spatial direction. However, the translational symmetry of the two-point function is manifest on the cylinder,
\begin{equation}\label{eq: XX-limiting-analysis}
    \braket{X^\mu(\tau_1,\sigma_1)X^\nu(\tau_2,\sigma_2)}= -\frac{c'}{2\epsilon}\log (2-2 \cos (\sigma_1-\sigma_2))\eta^{\mu\nu}\, .
\end{equation}
This reflects the fact that the intrinsic UR limit is defined on the cylinder. Another curious feature is the lack of time dependence in the correlator. The time independent nature of the two-point function is carried over to higher point functions as well. For example, a simple Wick's contraction procedure reveals the four-point function,\footnote{Odd-point functions of the Carroll scalar vanish due to the $X(\sigma,\t)\to-X(\sigma,\t)$ $\mathbb{Z}_2$ symmetry of the free theory.}
\begin{align}
    \braket{X^\mu(1)X^\nu (2) X^\rho(3)X^\sigma(4)}= -\frac{c'}{2\e} &\left[\log\left(2-\cos\sigma_{12}\right)\log\left(2-\cos \sigma_{34}\right)\eta^{\mu\nu}\eta^{\rho\sigma}\right.\nn\\
    &+\log\left(2-\cos\sigma_{13}\right)\log\left(2-\cos\sigma_{24}\right)\eta^{\mu\rho}\eta^{\nu\sigma}\nn\\
    &\left.+\log\left(2-\cos\sigma_{14}\right)\log\left(2-\cos\sigma_{23}\right)\eta^{\mu\sigma}\eta^{\nu\rho}\right]\, .
\end{align}
The above correlator can be easily obtained as a limit of the four-point function of the relativistic scalars in the UR limit, or by Wick's contraction procedure using \eqref{eq: XX-limiting-analysis}. As we show in the following sections, the time independence of the correlators plays a pivotal role in describing the nature of tensionless string theory in the induced representation.

\subsection{Naive high energy limit of primary correlators}

The map \eqref{eq: UR-limit-on-plane} plays an interesting role in the high energy limit of correlators of the vertex operators. We begin with the correlator of a general highest weight Virasoro primary,
\begin{equation}
    \braket{\mathcal{O}_{h_1,\bar h_1}(z_1,\bar z_1) \mathcal{O}_{h_2,\bar h_2}(z_2,\bar z_2)}=\frac{\delta_{h_1,h_2}\delta_{\bar h_1,\bar h_2}}{(z_{12})^{h_1+h_2} (\bar z_{12})^{\bar h_1+\bar h_2}}\, ,
\end{equation}

 The high energy limit on plane \eqref{eq:  UR-limit-on-plane},
\begin{equation}
     z_{12}\to x_{12}+\epsilon t_{12}\, ,\quad \bar z_{12}\to \frac{1}{x_{12}-\epsilon t_{12}}
\end{equation}
along with the replacement of Virasoro labels with the BMS labels,
\begin{equation}\label{eq: CFT2-weights-to-BMS-weights}
    h+\bar h=\frac{\xi}{\epsilon}\, ,\quad h-\bar h=\Delta\, ,
\end{equation}
gives the high energy limit of the correlator,
\begin{align}
    \braket{O_{h_1,\bar h_1}(1) O_{h_2,\bar h_2}(2)}
    &=x_{12}^{-2h}\left(1+\epsilon\frac{ t_{12}}{x_{12}}\right)^{-2h}x_{12}^{2\bar h}\left(1-\epsilon \frac{t_{12}}{x_{12}}\right)^{2\bar h}\, ,\nonumber\\
    &=x_{12}^{-2\Delta}\exp\left(-2\xi\frac{t_{12}}{x_{12}}\right)\, ,
\end{align}
where we have already imposed the constraint from the $\delta_{h_1,h_2}\delta_{\bar h_1,\bar h_2}$ by using $h_1=h_2=h$ and $\bar h_1=\bar h_2=\bar h$ in the first line.
In general, the two-point function of a CCFT$_2$ primary with weights,
\begin{equation}
    [L_0,\mathcal{O}_{\Delta,\xi}]=\Delta \mathcal{O}_{\Delta,\xi}\,,\quad [M_0,\mathcal{O}_{\Delta,\xi}]=\xi \mathcal{O}_{\Delta,\xi}
\end{equation}
obtained from the limit of the correlator of CFT$_2$ can be written as,
\begin{equation}\label{eq: CCFT-primary-correlator}
    \braket{O_{h_1,\bar h_1}(1) O_{h_2,\bar h_2}(2)}=\delta_{\Delta_1,\Delta_2} \delta_{\xi_1,\xi_2}x_{12}^{-2\Delta}\exp\left(-2\xi\frac{t_{12}}{x_{12}}\right)\, .
\end{equation}
Not surprisingly, we see that the correlators are the ones obtained from imposing Ward identities due to the global subalgebra of the BMS$_3$ algebra \cite{Bagchi:2009ca}. 

\medskip

Let us consider the case of the free boson again. The two-point relativistic CFT$_2$ current-current correlator is,
\begin{equation}
    \braket{\d_z X(1)\d_z X(2)}=-\frac{\alpha'}{2}\frac{1}{z_{12}^2}\, ,
\end{equation}
and the high energy limit of the correlator is,
\begin{equation}\label{eq: 2-pt-dXdX-plane}
    \braket{\d_{x} X(1)\d_x X(2)}=-\frac{c'}{2\e}\frac{1}{x_{12}^2}\, ,
\end{equation}
which can also be obtained by taking derivatives on the correlator \eqref{eq: 2pt-XX-plane}. The above correlator \eqref{eq: 2-pt-dXdX-plane} has the required translation  symmetry expected from a CCFT$_2$ primary. The term which breaks the spatial-translation symmetry in $\braket{XX}$ correlator vanishes in \eqref{eq: 2-pt-dXdX-plane}. The two-point function of the primary $\d_t X$, however, is trivial,
\begin{equation}
    \braket{\d_t X(1) \d_t X(2)}=0\, .
\end{equation}

\subsection{Vertex operators}

We wish to construct vertex operators for the null worldsheet in the induced vacuum. We do this by appealing to the tensile vertex operators and following their Carroll limit
\begin{equation}
    \mathcal{V}_p^{BMS}=\lim_{\epsilon\to 0}\mathcal{V}_p^{CFT}\, .
\end{equation}
The most subtle issue in the construction of these BMS vertex operators is that of normal ordering, which we postpone for the next section. We now specialise the discussion of the previous subsection to the BMS vertex operators. The two-point function of the Virasoro vertex operator is
\begin{equation}
    \braket{\mathcal{V}_{p_1}(z_1,\bar z_1) \mathcal{V}_{p_2}(z_2,\bar z_2)}=\frac{\delta_{p_1+p_2}}{(z_{12} \bar z_{12})^{-p_1 p_2 \alpha'/2}}\, ,
\end{equation}
since the conformal weights of the vertex operator $\mathcal{V}_p$ is, 
\begin{equation}
h=\bar h=p^2\alpha '/4\, .
\end{equation} 
The  leading two-point function of a BMS vertex operator obtained from the limit of the above CFT$_2$ correlator is
\begin{equation}\label{eq: VV-expectation-final}
     \braket{\mathcal{V}_{p_1}(x_1,t_1) \mathcal{V}_{p_2}(x_2,t_2)}=\delta_{p_1+p_2} x_{12}^{-2\Delta}\, ,
\end{equation}
where we have $\Delta\neq 0$ and $\xi=0$ contrary to a naive expectation with $\Delta=h-\bar h=0$, and $\xi=\epsilon(h+\bar h)=p^2 c'/2$, which appears for the highest weight representation in \cite{Hao:2021urq}. 

\medskip

In the next section we will show that the two-point function of the vertex operator that appears from an intrinsic definition  is indeed of the form \eqref{eq: VV-expectation-final}. In the process, we will also compute the value of $\Delta$ explicitly.

\section{Intrinsic Carroll analysis: Correlators and Vertex Operators }\label{sec: intrinsic-analysis}

The limiting analysis in the above section demonstrated the high energy behaviour of the correlators.  In this section, we provide an analysis intrinsic to tensionless strings. Precisely, we show how to obtain the expected correlators from the mode expansion of the free boson.

\subsection{Scalar two-point functions and OPE}
The limiting analysis two-point function of the scalar is invariant under spatial translation on the cylinder and completely independent of cylinder time $\tau$. 
We now calculate the two-point function of Carroll scalar in the induced vacuum $_I\braket {X^\mu X^\nu}_I$, using the mode expansion \eqref{eq: X-mode-expansion}.\footnote{From now on, we will skip the subscript $I$ on the induced vacuum.} The computation is straightforward till we obtain,
\begin{multline} \label{eq: XX-stop-1}
\langle X^\mu(1) X^\nu(2)\rangle=\braket{x^\mu x^\nu}+i\sqrt{\frac{c'}{2}}\sum_{n\neq 0}\frac{1}{n}\left(\braket{A^\mu_n x^\nu}e^{-in\sigma_1}+\braket{x^\mu A^\nu_n }e^{-in\sigma_2}\right)\\
-\frac{c'}{2}\sum_{\substack{n\\m\neq 0}}\frac{1}{nm}\braket{A_n^\mu A_m^\nu}e^{-in\sigma_1}e^{-im\sigma_2}\, ,
\end{multline}
where we have used the annihilation conditions on the right and the left induced vacuum,
\begin{equation}\label{eq: ind-annihilation-cond}
     B_n\ket{0}=0\, ,\quad \bra{0}B_n=0\, ,\quad \forall~ n\in\mathbb{Z}\,.
\end{equation}

We arrive at the condition on the left induced vacuum by imposing reality on the scalar field $X^\mu$. The reality condition $X^\mu=(X^\mu)^\dagger$ implies
\begin{equation}
    A^\dagger_n=A_{-n}\, ,\quad
    B^\dagger_n=B_{-n}\, .
\end{equation}

In \eqref{eq: XX-stop-1}, we ignore the constant piece $\braket{x^\mu x^\nu}$, and make the assumption
\begin{equation}\label{eq: XX-ass-1}
    \braket{A^\mu_n x^\nu}=\braket{x^\mu A^\nu_n }=0\, .
\end{equation}
This assumption is justified if we use the mode expansion $A_n=(1/\sqrt{\epsilon})(\alpha_n-\tilde\alpha_{-n})$ and use the fact that in the limit $\epsilon\to0$ the highest weight vacuum is the induced vacuum. Explicitly,
\begin{equation}
	\braket{A_n^\mu x^\nu}=\lim_{\epsilon\to0}\frac{1}{\sqrt{\epsilon}}\,{}_\alpha\braket{(\alpha_n^\mu-\tilde\alpha_{-n}^{\mu})x^\nu}_{\alpha}=0\, .
\end{equation}

The only remaining piece is the two point function $\braket{AA}$. 
To match the two-point function obtained from the limiting analysis \eqref{eq: XX-limiting-analysis} we find,\footnote{{Since the $A_n$ modes create a descendant state with zero mass but non-zero spin since $n\neq 0$. The descendant state is then orthogonal to the vacuum.}}
\begin{align}\label{eq: AA-2pt}
    \braket{A_n^\mu A_m^\nu}=\begin{cases}
                    m\delta_{n+m}\eta^{\mu\nu}\, , \quad n>0\, ,\\
                    n\delta_{n+m}\eta^{\mu\nu}\, ,\quad n<0\, ,
                     \end{cases}
\end{align}
along with our earlier assumption $\braket{x^\mu A_n^\nu}=\braket{A_n^\mu x^\nu }=0$ for all $n\in\mathbb{Z}_{\neq0}$.  We have carefully chosen the two-point function $\braket{A_n^\mu A_m^\nu}$ such that it is non-zero and is also consistent with the induced vacuum expectation value of the commutation relation,
\begin{align}
    \braket{[A_n^\mu\, , A_m^\nu]}&=\braket{A_n^\mu A_m^\nu}-\braket{ A_m^\nu A_n^\mu}=\begin{cases}
        m\delta_{n+m} \eta^{\mu\nu}-m\delta_{n+m} \eta^{\mu\nu}&=0 \quad \text{if } n>0\, ,m<0\, ,\\
        n\delta_{n+m} \eta^{\mu\nu}-n\delta_{n+m} \eta^{\mu\nu} &=0\quad\text{if } n<0\, ,m>0\, .
    \end{cases}
\end{align}
A compact expression for \eqref{eq: AA-2pt} is
\begin{equation}\label{eq: XX-ass-2}
    \braket{A_n^\mu A_m^\nu}=-|n|\delta_{n+m}\eta^{\mu\nu}\, .
\end{equation}
Again, we can motivate this choice from the limiting analysis on modes,
\begin{equation}\label{eq: AA-two-point-from-limit}
	 \braket{A_n^\mu A_m^\nu}=\lim_{\epsilon\to0}\frac{1}{\epsilon}\, {}_\alpha \braket{(\alpha_n^\mu-\tilde\alpha_{-n}^{\mu}) (\alpha_n^\nu-\tilde\alpha_{-n}^{\nu})}_{\alpha}=-|n|\delta_{n+m}\eta^{\mu\nu}\, .,
\end{equation}
where we have absorbed the factor of $1/\e$ in the normalisation. With the choices \eqref{eq: XX-ass-1}, \eqref{eq: XX-ass-2} we can proceed to compute the two-point function,
\begin{align}
    \langle X^\mu(1)X^\nu(2)\rangle&=-\frac{c'}{2}\sum_{\substack{n>0\\m\neq 0}}\frac{1}{nm}m\delta_{n+m}e^{-in\sigma_1}e^{-im\sigma_2}-\frac{c'}{2}\sum_{\substack{n<0\\m\neq 0}}\frac{1}{nm}n\delta_{n+m}e^{-in\sigma_1}e^{-im\sigma_2}\, ,\nn\\
    &=\frac{c'}{2}\sum_{n>0}\frac{1}{n}\cos(n\sigma_{12})\, .
\end{align}
The sum is,\footnote{This is a formal sum since the sum is not convergent.}
\begin{equation}\label{eq: XX-correlator-cylinder}
    \langle X^\mu(1)X^\nu(2)\rangle=-\frac{c'}{2}\log(2-2\cos(\sigma_{12}))\, ,
\end{equation}
which is spatially translation-invariant on the cylinder and time-independent.\footnote{{The two-point function $\langle X^\mu(1)X^\nu(2)\rangle$ matches with the limiting analysis upto an overall normalisation factor of $1/\epsilon$. The missing $\epsilon$ is the one we omitted in \eqref{eq: AA-two-point-from-limit}.}} Again, we use the cylinder to plane map \eqref{eq: cylinder-to-plane-coords} to find the two-point function on the plane,
\begin{equation}\label{eq: XX-correlator-plane}
     \langle X^\mu(1)X^\nu(2)\rangle=-\frac{c'}{2}\log\left(1-\frac{x_1}{x_2}\right)-\frac{c'}{2}\log\left(1-\frac{x_2}{x_1}\right) 
     =-\frac{c'}{2}\log\left(2-\frac{x_1}{x_2}-\frac{x_2}{x_1}\right)\,.
\end{equation}
Notice that the two-point function loses the translation invariance after switching to the plane coordinates.

\subsection{Multiplet structure}\label{sec: multiplet structure}
The two-point function of the scalar $\langle X^\mu X^\nu\rangle$ is not the two-point function of a primary operator, as it does not satisfy conformal Carroll Ward identities. The non-primary nature of the Carroll scalar $X$ is also expected from a limit of CFT$_2$. Rather than the scalar itself, we expect the derivatives of the scalar operator $\partial_x X^\mu$ and $\partial_t X^\mu$ to be the primary operators.\footnote{We can similarly consider the primaries on the cylinder, $\d_\sigma X$ and $\d_\t X$.} The mode expansions of the Carroll primaries are,
\begin{subequations}
  \begin{align}
	\d_x X^\mu&=i\sqrt{\frac{c'}{2}}\sum_{n\in\mathbb{Z}}(n+1)B^\mu_n t x^{-n-2}-i\sqrt{\frac{c'}{2}}\sum_{n\neq0}A^\mu_nx^{-n-1}\, ,\\
	\d_t X^\mu&=-i\sqrt{\frac{c'}{2}}\sum_{n\in\mathbb{Z}}B^\mu_n x^{-n-1}\, .
\end{align}  
\end{subequations}

The weights obtained in the previous section from the high energy limit of the correlators can be verified by computing the commutators explicitly,
\begin{align}
	[L_n,\d_x X^\mu]&=i\frac{\sqrt {c'}}{2\sqrt{2}}\left[\sum_{m\in\mathbb{Z}}A_{n-m}\cdot B_m,\sum_{l\in\mathbb{Z}}(l+1)B^\mu_l t x^{-l-2}-\sum_{l\neq0}A^\mu_l x^{-l-1}\right]\, ,\nonumber\\
		&=(x^{n+1}\d_x+ (n+1)x^n +(n+1)x^n t\d_t)\d_x X^\mu-n(n+1)x^{n-1}t\d_tX^\mu,
\end{align}
and,
    \begin{align}
        [L_n,\d_t X^\mu]=&-i\frac{\sqrt {c'}}{2\sqrt{2}}\sum_{\substack{m\in\mathbb{Z}\\l\in\mathbb{Z}}}\left[A_{n-m}\cdot B_m,B^\mu_l x^{-l-1}\right]\nn\\
        &=(x^{n+1}\d_x+(n+1)x^n)\d_t X(x,t)\, .
        \end{align}
That is, the above commutation relations are exactly the same as the highest weight commutators with $\Delta=1$ for both $\d_x X(x,t)$ and $\d_t X(x,t)$. There is, however, a non-trivial $\xi$ which is a $2\times 2$ matrix,
\begin{equation}
    \xi=\begin{pmatrix}
        0 & 0\\
        1 & 0
    \end{pmatrix}\,.
\end{equation}
To verify the structure of $\xi$ matrix clearly visible, we calculate
\begin{align}
	[M_n,\d_x X^\mu]&=i\frac{\sqrt {c'}}{2\sqrt{2}}\left[\sum_{m\in\mathbb{Z}}B_{n-m}B_m,\sum_{l\in\mathbb{Z}}(l+1)B_l t x^{-l-2}-\sum_{l\neq0}A_l x^{-l-1}\right]\, ,\nonumber\\
    &=x^{n+1}\d_t(\d_x X)+x^n(n+1)(\d_t X)
\end{align}
and the trivial commutator,
\begin{equation}
    [M_n,\d_t X^\mu]=0\, .
\end{equation}

The multiplet structure was previously discussed for the highest weight representation of the 2d Carroll scalar in \cite{Hao:2021urq}. The multiplet structure is akin to a $\log$ partner in a relativistic $\log$ CFT$_2$ \cite{Gurarie:1993xq} \footnote{See \cite{Bagchi:2009pe} for the first mentions of a connection between Log CFTs and non-Lorentzian CFTs.}. The explicit multiplet structure thus obtained is
\begin{equation}
    M_0\begin{pmatrix}
        \partial_t X\\
        \partial_x X
    \end{pmatrix}=
    \begin{pmatrix}
        0\\
        \partial_t X
    \end{pmatrix}\, .
\end{equation}
Since the two-point function $\langle X^\mu(1)X^\nu(2)\rangle$ is independent of time we have the two-point functions from the mode expansion as,
\begin{subequations}\label{eq: der-X-2pt-plane}
 \begin{align}
    \partial_{x_1}\partial_{x_2}\langle X^\mu(1)X^\nu(2)\rangle&=-\frac{c'}{x_{12}^2}\, ,\\
    \partial_{t_1}\partial_{x_2}\langle X^\mu(1)X^\nu(2)\rangle&=\partial_{t_1}\partial_{t_2}\langle X^\mu(1)X^\nu(2)\rangle=0\, .  
\end{align}   
\end{subequations}
Or in cylinder coordinates, 
\begin{subequations}\label{eq: der-X-2pt-cylinder}
   \begin{align}
    \partial_{\sigma_1}\partial_{\sigma_2}\langle X^\mu(1)X^\nu(2)\rangle&=-\frac{c'}{2-2\cos{\sigma_{12}}}\, ,\\
    \partial_{\tau_1}\partial_{\sigma_2}\langle X^\mu(1)X^\nu(2)\rangle&=\partial_{\tau_1}\partial_{\tau_2}\langle X^\mu(1)X^\nu(2)\rangle=0\, . 
\end{align} 
\end{subequations}
In the OPE limit, $\sigma_{1}\sim\sigma_2$, we have the non-trivial two-point function,
\begin{equation}
    \partial_{\sigma_1}\partial_{\sigma_2}\langle X^\mu(1)X^\nu(2)\rangle=-\frac{c'}{\sigma_{12}^2}\,,
\end{equation}
as expected. 

\subsection{Vertex operators and correlators}
The vertex operators in the tensionless string theory are defined as
\begin{equation}
    \mathcal{V}_p(\sigma,\tau)=:\exp(i p\cdot X(\sigma,\tau)):\, ,
\end{equation}
where $X(\sigma,\tau)$ is the Carroll boson defined on the cylinder. The cylinder mode expansion of the free Carroll boson is \eqref{eq: X-mode-expansion}. There are many choices of normal ordering of the vertex operator in tensionless string theory. A simple choice is to use the $AB$ normal ordering, i.e., the $B_n$ modes are pushed to the right,\begin{equation}
:A_nB_m:=A_nB_m\, ,\quad  \forall n,m\in\mathbb{Z}\, .
\end{equation}
However, with $AB$ normal ordering we get correlators with various complications (see appendix \ref{app: AB-normal-order}), and we will look for another prescription \footnote{Weyl ordering of  $A, B$ oscillators also has complications and hence we ignore it.}. We will be inspired by the limit from usual string theory.

\medskip

Tensile oscillators $\alpha$ and $\tilde \alpha$ of course have a well-defined normal ordering for the highest weight representation in CFT$_2$. We choose the same normal ordering,
\begin{equation}\label{eq:alpha-normal-ordering}
    :\alpha_n\alpha_m:=\alpha_n\alpha_m\, ,\quad \text{if } m\geq n\, ,
\end{equation}
and follow the limit for the tensionless string. 
The 2d Carroll boson $X^\mu$ can be written in terms of $\alpha$ oscillators using the transformation \eqref{eq: AB-to-alphas}.
The vertex operator can be expanded in the $\alpha$-normal ordering,
\begin{multline}    \mathcal{V}_{p}=\exp{\left\{i\sqrt{\frac{c'}{2}}p\cdot\left(\sqrt{\frac{2}{c'}}\phi_0+\sum_{n>0}\Big(\tilde \alpha_{-n}e^{-in\sigma}+\alpha_{-n}e^{in\sigma}\Big)\left(\sqrt{\epsilon} \tau-\frac{i}{n\sqrt{\epsilon}}\right)\right)\right\}}\\
\times\exp\left\{i\sqrt{\frac{c'}{2}}p\cdot\left(\tau\sqrt{\epsilon}(\alpha_0+\tilde\alpha_0)+\sum_{n>0}\Big(\tilde \alpha_{n}e^{in\sigma}+\alpha_{n}e^{-in\sigma}\Big)\left(\sqrt{\epsilon} \tau+\frac{i}{n\sqrt{\epsilon}}\right)\right)\right\}\, .
\end{multline}
We use the BCH formula to bring the oscillators to a normal ordered form. Schematically, the calculation proceeds as,
\begin{align}
:\mathcal{V}_{p_1}::\mathcal{V}_{p_2}:&=e^{ip_1\times\text{Creation modes}}\underbrace{e^{ip_1\times\text{Annihilation modes}}e^{ip_2\times\text{Creation modes}}}_{\text{apply BCH formula}}e^{ip_2\times\text{Annihilation modes}}\, ,
\end{align}
where we use the commutation relation between the annihilation ($A$) and creation ($C$) modes\footnote{$A(1)$ denotes the annihilation modes inserted at coordinate $(t_1,x_1)$ while $C(2)$ denotes the creation modes inserted at coordinate $(t_2,x_2)$.},
\begin{align}
    [A(1),C(2)]= -\frac{c'}{2\epsilon}\log\left({2-2\cos{\sigma_{12}}}\right)\, ,
\end{align}
to finally obtain the product,
\begin{equation}
    \mathcal{V}_{p_1}(\tau_1,\sigma_1)\mathcal{V}_{p_2}(\tau_2,\sigma_2)\sim\mathcal{V}_{p_1+p_2}(2-2\cos(\sigma_1-\sigma_2))^{\p_1\p_2\frac{c'}{2}}\, ,
\end{equation}
where we have introduced the rescaled momenta $\p=p/\sqrt\e$. The two-point function is,
\begin{equation}
    \braket{\mathcal{V}_{p_1}(1)\mathcal{V}_{p_2}(2)}=\delta(\p_1+\p_2)(2-2\cos(\sigma_1-\sigma_2))^{\p_1\p_2\frac{c'}{2}}\, .
\end{equation}
where we now have a momentum conserving $\delta_{\p_1+\p_2}$. The above correlator is translation invariant and time independent and satisfies the Ward identities imposed by the global BMS$_3$ generators which simply reduces to the Ward identities due to a single copy of the Virasoro algebra with,
\begin{equation}\label{eq: delta-xi-Vp}
    \Delta=\frac{c'\p^2}{2}\, ,\quad \xi=0\, .
\end{equation}
Thus the induced representation vertex operator is at least a BMS quasi-primary. In appendix \ref{app: BMS-weights} we show the computation of the eigenweights by directly computing the commutators of the BMS generators with the vertex operators. The cylinder to plane map \eqref{eq: cylinder-to-plane-coords} connects the above cylinder two-point function to the plane two-point function, 
\begin{equation}\label{eq:def of V op}
 \mathcal{V}_p(\sigma,\tau)=x^\Delta e^{i\tau\xi}\mathcal{V}_{p}(x,t)\,.
\end{equation}

Using $\Delta=\p^2 c'/2$ and $\xi=0$ we get the familiar two-point correlator on the plane,
\begin{align}
	\braket{\mathcal{V}_p(x_1, t_1) \mathcal{V}_p(x_2,t_2)}(x_1x_2)^{\Delta}&=(-x_{12}^2)^{\p_1 \p_2 c'/2} (x_1 x_2)^{-\p_1 \p_2 c'/2}\delta(\p_1+\p_2)\, ,\nonumber\\
	\braket{\mathcal{V}_p(x_1, t_1) \mathcal{V}_p(x_2,t_2)}&=(-x_{12}^2)^{\p_1\p_2 c'/2}\delta(\p_1+\p_2)\, .
\end{align}
In particular, we find that the two-point function is also translational invariant on the plane. We can proceed exactly as above to compute higher point correlators of the vertex operators.

\subsection{Higher point functions of vertex operators}

The product of three vertex operators can be deduced similarly,
\begin{equation}
	\mathcal{V}_{p_1}(1)\mathcal{V}_{p_2}(2)\mathcal{V}_{p_3}(3)= \prod_{j=1}^{3}:\exp\left( ip_j\cdot X(j)\right):\, .
\end{equation}
We again bring the creation modes from all vertex operators to the left and the annihilation operators to the right, which gives the product,
\begin{align}\label{3pt-vertices-line1}
	\mathcal{V}_{p_1}\mathcal{V}_{p_2}\mathcal{V}_{p_3}&=:e^{C(1)}e^{A(1)}e^{C(2)}e^{A(2)}e^{C(3)}e^{A(3)}:\, ,\nonumber\\
	&=e^{C(1)}e^{C(2)}e^{C(3)}e^{A(1)}e^{A(2)}e^{A(3)}e^{[A(1), C(2)]}e^{[A(2), C(3)]}e^{[A(1), C(3)]}\, ,
\end{align}
where we denote the creation oscillators as $C(i)$ and the annihilation oscillators as $A(i)$.\footnote{The creation and annihilation operators, $C(i)$ and $A(i)$, already contain the multiplicative factor of $i p_i$.} The commutators are,
\begin{equation}\label{eq: [A,C]}
	[A(i), C(j)]=\,\p_i \p_j\frac{c'}{2}\log(2-2\cos{\sigma_{ij})}\, .
\end{equation}
Using the above commutator \eqref{eq: [A,C]} in the previous equation \eqref{3pt-vertices-line1} we get,
\begin{equation}
\mathcal{V}_{p_1}\mathcal{V}_{p_2}\mathcal{V}_{p_3}\sim \mathcal{V}_{p_1+p_2+p_3}\prod_{j>i=1}^3\left(2-2\cos{\sigma_{ij}}\right)^{c'\p_i\p_j/2}\, .
\end{equation}

The above answer can be generalised to any arbitrary $n$-point function of the vertex operators, and can be shown to be true by Wick's contraction.
\begin{equation}\label{Vertex operator n-point function}
		\mathcal{V}_{p_1}\mathcal{V}_{p_2}\cdots\mathcal{V}_{p_n}=\exp{\left(\sum_{i=1}^{n}C(i)\right)}\exp{\left(\sum_{i=1}^{n}A(i)\right)}\prod_{i<j}(2-2\cos{\sigma_{ij}})^{\p_i\p_j c'/2}\, 
\end{equation}
where $i,j=1,\cdots,n$ in the product. The three-point function on the plane is then,
\begin{equation}\label{eq: three-point-vertexes}
\braket{\mathcal{V}_{p_1}\mathcal{V}_{p_2}\mathcal{V}_{p_3}}=\left(x_{12}^2\right)^{c'\p_1\p_2/2}\left(x_{23}^2\right)^{c'\p_2\p_3/2}\left(x_{13}^2\right)^{c'\p_1\p_3/2}\delta(\p_1+\p_2+\p_3)\, ,
\end{equation}
where the momentum conservation of the three-point function is imposed. Similarly, we can write the expression of any $n$-point function, 
	\begin{equation}\label{eq: n-point-vertexes}
		\braket{\mathcal{V}_{p_1}(1)\mathcal{V}_{p_2}(2)\cdots\mathcal{V}_{p_n}(n)}=\prod_{j>i=1}^n(x_{ij}^2)^{c'\p_i\p_j/2}\delta(\p_1+\cdots+\p_n)\,.
	\end{equation}
 The $n$-point function resembles the $n$-point function of vertex operators used to calculate scattering amplitudes in open string theory with the circle coordinate $x$ taking the place of the boundary coordinate. Thus, the correlators \eqref{eq: n-point-vertexes} which are functions of only the spatial coordinates hints at the emergence of tensionless open strings from closed strings.

\section{Scattering amplitude in tensionless string theory}\label{sec: scattering}

We are interested in the very high energy limit of critical (tensile) string theory. This high energy sector was previously accessed by constructing scattering amplitudes with a finite tension and looking at the high energy limit of these tensile string amplitudes \cite{Gross:1987kza}. Since string amplitudes are analytic functions of $\alpha'$, they can be expanded in the large $\alpha'$ limit. We have stressed that tensionless strings on the null worldsheet in the induced vacuum putatively describe the ultra-high energy limit of string theory. 

In this section,  we build scattering amplitudes for tensionless strings from the worldsheet methods outlined in the previous sections and obtain the high energy amplitudes of Gross-Mende \cite{Gross:1987kza}. We find that these tensionless amplitudes are exactly the same as the high energy limit of tensile string amplitudes.

\subsection{Integrated vertex operator}

The scattering amplitudes of tensionless string theory should be invariant under worldsheet BMS diffeomorphisms, since the amplitude in the target space is independent of the positions of the vertex operators. The leading correlator of the vertex operators is agnostic of the time coordinates of the insertions since the temporal direction becomes null in the tensionless limit. Thus, the distances along the null directions are meaningless at least in the leading order. If we treat the circle coordinate $x$ as the boundary coordinate, then the correlators look similar to the boundary correlators of open string theory, but with large momenta.

\medskip

In close analogy with the integrated vertex operator of open string theory on a flat background, we can define the integrated vertex operator of the tensionless string as,
\begin{equation}\label{intvert}
	V_p=\tilde{g}\int dx\,\mathcal{V}_p(\xi_i) \,,
\end{equation}
where $\tilde{g}$ is a parameter analogous to the tensile string coupling. Usually, open tensile string vertex operators are defined with the open string coupling $g_o$ and closed ones with the closed string coupling $g_s$. Since there is a blurring of closed and open strings in the tensionless limit, we don't label our coupling here with $o$ or $s$. At first sight the vertex operator \eqref{intvert} looks like an open string analogue. But we stress that this holds for both open and closed null strings. The reason behind not integrating over the null direction $\tau$ is that it would not achieve much since the correlation functions of these null string vertex operators are independent of $\tau$. \footnote{In the appendix \ref{app: closed-type-integrated-Vp}, we present the vertex operator integrated over both the worldsheet coordinates in close analogy with the closed string vertex operator. However, the closed-type tensionless integrated vertex operator results in spurious poles in the four-point amplitude. The lightest pole in the closed-type amplitude corresponds to a state which seems to be absent in the spectrum \eqref{eq: closed-mass-spectrum}. We don't understand the significance of this spurious pole and will conclude from this exercise that the genuine integrated null string vertex operator is the one we have defined in \eqref{intvert}.}

\medskip

The worldsheet diffeomorphism invariance of the scattering amplitude requires the integrated vertex operator to be invariant under the global BMS transformations.\footnote{Physically, the integrated vertex operator represents a scattering state in target spacetime and is independent of the worldsheet coordinates.} The integration measure transforms under a general BMS coordinate transformation,
\begin{equation}
    x\to x'=f(x)\, ,\quad t\to t'= tf'(x)+g(x)
\end{equation}
as,
\begin{equation}
    dx'=|f'(x)| dx\, .
\end{equation}
An important observation is that the integration measure is invariant under supertranslations. In particular, the measure does not transform for a trivial $f(x)$ and any arbitrary choice of non-trivial $g(x)$. The agnostic behaviour of the measure under supertranslations can also be seen from the limit (see appendix \ref{app: measure}).
The diffeomorphism invariance of $V_p$ gives
\begin{equation}
    \Delta=1\, ,\quad \xi=0\, .
\end{equation}
which implies for the rescaled momenta,
\begin{equation}
    \p^2=\frac{2}{c'}\,.
\end{equation}
In terms of the physical momenta, we have the on-shell scattering state with
\begin{equation}
  p^2=\frac{2\epsilon}{c'}\, ,
\end{equation}
where $\lim \epsilon\to 0$ essentially implies that the state is massless. We see that the tachyon state of tensile string theory becomes massless in the tensionless limit. The massless spectra in the theory built on the induced vacuum is expected since the tensionless theory is the high energy limit of string theory.

\subsection{Null string scattering amplitudes}
We begin with the computation of the tree-level three point scattering amplitude in tensionless strings and then we move to the non-trivial tree-level four-point scattering amplitude.

\subsubsection*{Three-point scattering amplitude}
The three-point amplitude of the tachyon-like vertex operators is,\footnote{Note that in the induced representation of the bosonic tensionless string theory we do not have a fixed dimension of the target space \cite{Bagchi:2020fpr}.}
\begin{equation}\label{eq:three-point amplitude o}
    \mathcal{A}^{(3)}(p_1,p_2,p_3)=\frac{(2\pi)^D\tilde{g}^{3/2}}{\text{Vol~(ISO(2,1))}}\int\prod_{i=1}^3dx_i\braket{\mathcal{V}_{p_1}(\xi_1)\mathcal{V}_{p_2}(\xi_2)\mathcal{V}_{p_3}(\xi_3)}\, ,
\end{equation}
where we have introduced a coupling $\tilde g$. The three-point function  of the vertex operators is \eqref{eq: three-point-vertexes},
\begin{equation}
    \braket{\mathcal{V}_{p_1}(1)\mathcal{V}_{p_2}(2)\mathcal{V}_{p_3}(3)}=(x_{12}^2)^{\p_1\p_2c'/2}(x_{23}^2)^{\p_2\p_3c'/2}(x_{13}^2)^{\p_1\p_3c'/2}\delta(\p_1+\p_2+\p_3)\,. 
\end{equation}
The surviving global part of the gauge symmetry on the plane,
\begin{equation}
    x\to x'=\frac{ax+b}{cx+d}\, ,\quad t\to t'=\frac{t}{(cx+d)^2}\, ,
\end{equation}
with $ad-bc=1$, allows us to fix the positions of the three vertex operators to
\begin{equation}
    (t_1,x_1)=(0,0)\,,~~~(t_2,x_2)=(0,1)\,,~~~(t_3,x_3)=(0,\infty)\,.
\end{equation}
Therefore, the three-point correlator is given as,
\begin{align}
\braket{\mathcal{V}_{p_1}(0,0)\mathcal{V}_{p_2}(0,1)\mathcal{V}_{p_3}(0,\infty)}&=\lim_{x_3\to\infty}x_3^{\p_3^2c'}(1-x_3)^{\p_2\p_3c'}(-x_{3})^{\p_1\p_3c'}\delta(\p_1+\p_2+\p_3)\,,\nonumber\\
&=(-1)^{-\p_3^2c'}\delta(\p_1+\p_2+\p_3)\,,
\end{align}
The on-shell condition of the physical momentum $p_3^2 c'/\e=2$ implies,
\begin{equation}
    \braket{\mathcal{V}_{p_1}(0,0)\mathcal{V}_{p_2}(0,1)\mathcal{V}_{p_3}(0,\infty)}=\delta(\p_1+\p_2+\p_3)\, .
\end{equation}
Substituting the three-point correlator along with the positions of the vertex operators in \eqref{eq:three-point amplitude o}, we get
\begin{equation}
        \mathcal{A}^{(3)}(p_1,p_2,p_3)=(2\pi)^D\tilde{g}^{3/2}\delta(\p_1+\p_2+\p_3)\,.
\end{equation}
The global part of the residual gauge symmetry completely fixes the three-tachyon tree-level scattering amplitude. We now proceed to the four-point scattering amplitude.

\subsubsection*{Four-point amplitude}
Similar to the three-point tachyon-like amplitude, the 2-2 scattering amplitude of tachyon-like vertex operators is,
\begin{equation}\label{eq:four-point amplitude1o}
\mathcal{A}^{(4)}_s(p_1,p_2,p_3,p_4)=\frac{(2\pi)^D\tilde{g}^2}{\text{Vol~(ISO(2,1))}}\int\prod_{i=1}^4dx_i\braket{\mathcal{V}_{p_1}(x_1,t_1)\dots\mathcal{V}_{p_4}(x_4,t_4)}\, .
\end{equation}

The four-point function of the vertex operators is \eqref{eq: n-point-vertexes},
\begin{equation}\label{eq: four-point correlator 1o}
	\braket{\mathcal{V}_{p_1}(1)\cdots\mathcal{V}_{p_4}(4)}=\delta\left(\sum_{i=1}^{4}p_i\right)\prod_{i<j=1}^{4}(x_{ij}^2)^{\p_i\p_j c'/2}\, ,
\end{equation}
where we have introduced rescaled momenta $\p_i=p_i/\sqrt{\e}$. We can again use the residual global symmetry $ISO(2,1)$ to fix the positions of three of the four vertex operators. We not only need to choose three out of the four positions for the vertex operators but also order them accordingly. One of the possible coordinate choices of the $x$-ordered vertex operators is
\begin{equation}\label{eq: x ordered vertexop}
(t_1,x_1)=(0,0)\,,\quad(t_2,x_2)=(t,x)\,,\quad(t_3,x_3)=(0,1)\,,\quad (t_4,x_4)=(0,\infty)\,,
\end{equation}
where $0<t,x<1$. The four point correlator \eqref{eq: four-point correlator 1o} for these positions of the vertex operators reduces to,
\begin{align}\label{eq: four-point correlator 2o}
    \braket{\mathcal{V}_{p_1}(1)\cdots\mathcal{V}_{p_4}(4)}&=\delta\left(\sum_{i=1}^{4}p_i\right)(-1)^{-\p_1^2c'-\p_2\p_1 c'-\p_2^2 c'+\p_3\p_4 c'}(x)^{\p_1\p_2 c'}(1-x)^{\p_2\p_3 c'}\,,\nonumber\\
    &=\delta\left(\sum_{i=1}^{4}p_i\right)x^{-2-s/2}(1-x)^{-2-u/2}\,.
\end{align}
The above represents one of the partial amplitudes. The total amplitude is the sum of the $s\, ,t\, ,u$-channels, where we have defined the Mandelstam variables with the tensionless scale $c'$ and the rescaled momenta $\p_i$,
\begin{equation}\label{Mandelstam variables}
    s=-c'(\p_1+\p_2)^2\,,\quad
    t=-c'(\p_1+\p_3)^2\,,\quad
    u=-c'(\p_1+\p_4)^2\,,\quad s+t+u=-8\,.
\end{equation}
\footnote{We use the unprimed notation for the rescaled Mandelstam variables, since these are exactly the Mandelstam variables of string theory
\begin{equation*}
    s=-\alpha'(p_1+p_2)^2=-\frac{c'}{\e}(p_1+p_2)^2=-c'(\p_1+\p_2)^2\, .
\end{equation*}}
The total amplitude is the sum of the $s\, ,t\, ,u$-channels.

The Mandelstam variables are large as they scale as $1/\e$ for finite $p_i$. However, we will continue with the computation without utilising the largeness of the Mandelstam variables. Only at the final step of the computation will we utilise the condition that the Mandelstam variables are large as compared to the mass of tensionless states. Remember that the mass shell condition for the vertex operators in terms of the rescaled momenta is $(\p_i)^2=2/c'$. 
\medskip

The four-point amplitude in \eqref{eq:four-point amplitude1o} can be computed by substituting \eqref{eq: four-point correlator 2o} in \eqref{eq:four-point amplitude1o}, we get
\begin{equation}\label{eq:four-point amplitude2o}
\mathcal{A}^{(4)}(p_1,p_2,p_3,p_4)=(2\pi)^D\tilde{g}^2\delta\left(\sum_{i=1}^{4}p_i\right)\underbrace{\int_0^1dx~x^{-2-s/2}(1-x)^{-2-u/2}}_{I(s,u)}\,.
\end{equation}
The integral converges in the region $\Re(s)<-2$ and $\Re(u)<-2$. As $s\to-2$, the integral diverges at $x=0$. The divergence can be analysed by approximating the integrand $I(s,u)$ in a neighbourhood of $x=0$
\begin{subequations}
    \begin{align}
     I(s,u)&=\int_0^r dx ~x^{-2-s/2}~+\text{terms analytic at}~s=-2\,,\\
     &=-\frac{r^{-1-s/2}}{s/2-1}~+\text{terms analytic at}~s=-2\,.
\end{align} 
\end{subequations}
Therefore, the s-channel pole is at $s=-2$. The integral \eqref{eq:four-point amplitude2o} evaluates to
\begin{equation}
    \mathcal{A}^{(4)}(p_1,p_2,p_3,p_4)=(2\pi)^D\tilde{g}^2\delta\left(\sum_{i=1}^{4}p_i\right)\frac{\Gamma(-1-s/2)\Gamma(-1-u/2)}{\Gamma(2+t/2)}\,,
\end{equation}
with $\Re(s)\,,~$$\Re(u)<-2$. Adding all the partial amplitudes, we obtain the full crossing-symmetric amplitude,
\begin{multline}\label{eq:four-point amplitude3o}
    \mathcal{A}^{(4)}(p_1,p_2,p_3,p_4)=(2\pi)^D\tilde{g}^2\delta\left(\sum_{i=1}^{4}p_i\right)\Bigg(\frac{\Gamma(-1-s/2)\Gamma(-1-u/2)}{\Gamma(2+t/2)}\\
    +\frac{\Gamma(-1-s/2)\Gamma(-1-t/2)}{\Gamma(2+u/2)}+\frac{\Gamma(-1-t/2)\Gamma(-1-u/2)}{\Gamma(2+s/2)}\Bigg)\,.
\end{multline}
The amplitude is similar to the \textit{Veneziano amplitude} where we have $s/2\, ,t/2\, ,u/2$ instead of $s,t,u$. In string theory, there is a doubling of the Mandelstam variables due to the doubling trick, where we interpret the closed string as a double of the open string. 

\medskip

To access the physical domain of the Mandelstam variables, that is, $s>0$ and $t,u<0$, we analytically continue the amplitude \eqref{eq:four-point amplitude3o},
\begin{multline}
   \mathcal{A}^{(4)}(p_1,p_2,p_3,p_4)=(2\pi)^D\tilde{g}^2\delta\left(\sum_{i=1}^{4}p_i\right)\frac{\Gamma(-t/2)\Gamma(-u/2)}{\Gamma(s/2)}\frac{16}{s(s+2)(t+2)(u+2)}\\\times\Bigg(\frac{\sin(\pi t/2)}{\sin(\pi s/2)}+\frac{\sin(\pi u/2)}{\sin(\pi s/2)}+1\Bigg)\,. 
\end{multline}

\begin{figure}
   \subfloat[s-channel \label{fig:PKR}]{%
      \includegraphics[ width=0.32\textwidth]{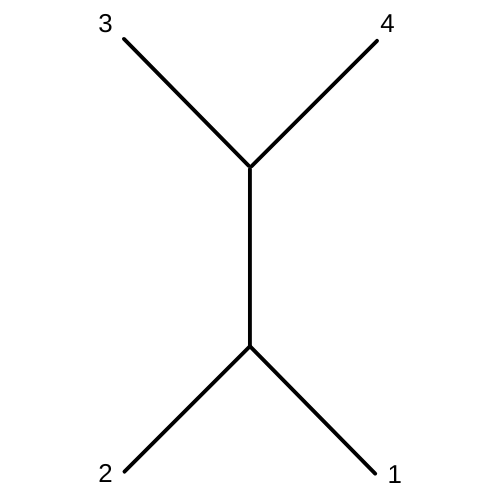}}
\hspace{\fill}
   \subfloat[t-channel \label{fig:PKT} ]{%
      \includegraphics[ width=0.32\textwidth]{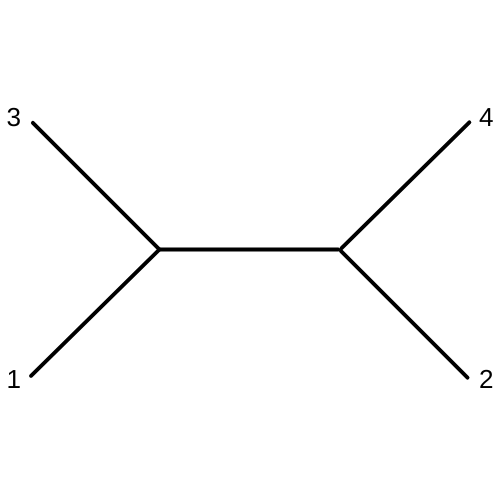}}
\hspace{\fill}
   \subfloat[u-channel \label{fig:tie5}]{%
      \includegraphics[ width=0.32\textwidth]{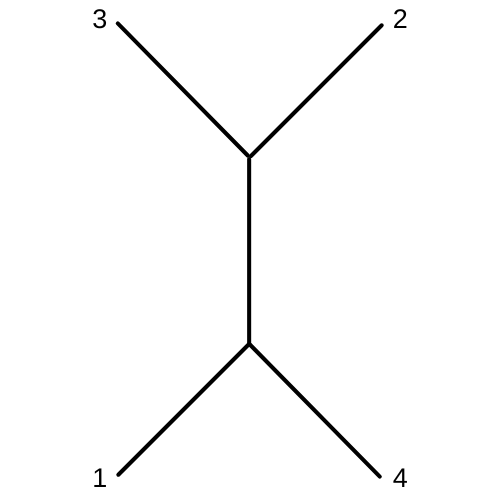}}\\
\caption{Different scattering channels.}
    \label{workflow}
\end{figure}

The pole structure is now clearly visible. The poles are located at the values $s=2n$, with $n\in\mathbb{Z}_{\geq -1}$. The first pole matches with the \textit{tachyon} state with $m'^2=-2/c'$. The remaining poles correspond to the higher excited states,
\begin{equation}\label{eq: open-mass-spectrum}
    m'^2=\frac{2n}{c'}\,\quad n\in\mathbb{Z}_{\geq 0} .
\end{equation}
In other words, the rescaled mass-shell condition is expected to have a tachyon state, a massless state, and higher excited states in the spectrum. 
\medskip

We can analyse the leading behaviour and the fluctuations around the leading behaviour of the amplitude by using Stirling's formula \eqref{eq:Stirling's Formula},
\begin{align}
    \frac{\Gamma(-t/2)\Gamma(-u/2)}{\Gamma(s/2)}&=-(2\pi)^{1/2} (-1)^{-t/2-u/2}e^{s/2+t/2+u/2}\left(\frac{2s}{tu}\right)^{1/2}t^{-t/2} u^{-u/2} s^{-s/2}\nonumber\\
    &\quad\times\exp\left[-\sum_{k=1}^\infty\frac{\zeta(1-2k)}{2k-1}\left(\left(\frac{t}{2}\right)^{1-2k}+\left(\frac{u}{2}\right)^{1-2k}+\left(\frac{s}{2}\right)^{1-2k}\right)\right]\, \nonumber\\
    &=-2\pi^{1/2}e^{-4}(-1)^{s/2} \left(\frac{s}{tu}\right)^{1/2}t^{-t/2} u^{-u/2} s^{-s/2}\,\nonumber\\
    &\quad\times\exp\left[-\sum_{k=1}^\infty\frac{2^{2k-1}\zeta(1-2k)}{2k-1}\left(t^{1-2k}+u^{1-2k}+s^{1-2k}\right)\right]\,,
\end{align}
where we have used the mass-shell condition $s+t+u=-8$. At this stage, however, the analysis still allows for finite-angle as well as infinitesimal scattering angle $\theta$ in the centre of mass frame. In other words, we can obtain both the Gross-Mende and the Regge behaviour of the amplitude \eqref{eq:four-point amplitude3o}. \\

\emph{Hard scattering limit}: The limit on the four-point scattering amplitude $s\to\infty$, and $t,u\to -\infty$ while keeping $s/t$ fixed is the hard-scattering limit of the tensile string theory studied in \cite{Gross:1987kza}. Finally, we now recognise the large nature of the  Mandelstam variables and find the asymptotic behaviour of the amplitude,
\begin{equation}\label{eq: 4-amplitude -asymptotic-form-1}
   \mathcal{A}^{(4)}(p_i)\sim(-1)^{s/2+1}\frac{16}{(stu)^{3/2}}\Bigg(\frac{\sin(\pi t/2)}{\sin(\pi s/2)}+\frac{\sin(\pi u/2)}{\sin(\pi s/2)}+1\Bigg)t^{-t/2} u^{-u/2} s^{-s/2}\,.
\end{equation}

In the\textit{ centre of momentum frame} we can choose the momenta,
\begin{equation}\label{eq:COM frame}
    \p_1=\frac{1}{2}(E',\vec{k}')\,,~~\p_2=\frac{1}{2}(E',-\vec{k}')\,,~~\p_3=\frac{1}{2}(-E',\vec{\tilde{k}}')\,,~~\p_4=\frac{1}{2}(-E',-\vec{\tilde{k}}')\,,
\end{equation}
where $E'$ is the total energy in the  centre of mass frame of incoming tachyons with momenta $\p_1$ and $\p_2$ with the scattering angle $\theta$. The Mandelstam variables in the centre of momentum frame take the form,

\begin{equation}\label{eq:s-t-u-com-frameo}
    s=c'E'^2\, ,\quad t=-(8+c'E'^2)\cos^2\frac{\theta}{2}\, ,\quad u=-(8+c'E'^2)\sin^2\frac{\theta}{2}
\end{equation}
Then the hard scattering limit of the four-point scattering amplitude can be expressed as,
\begin{equation}
    \mathcal{A}^{(4)}(\{p_i\})\sim (-1)^{s/2} \exp\left(-\frac{s}{2}f(\theta)\right)\, ,
\end{equation}
where,
\begin{equation}
    f(\theta)=-\sin^2\frac{\theta}{2}\ln{\sin^2\frac{\theta}{2}}-\cos^2\frac{\theta}{2}\ln{\cos^2\frac{\theta}{2}}\, .
\end{equation}

\emph{Regge limit}: in this limit, $s\to\infty$, while $u$ is kept fixed. It is still possible to keep the variable $u$ finite even as it naively scales like $1/\e$. Explicitly,
\begin{equation}
    u=-c'(\p_1+\p_4)^2=-4-2c'\frac{p_1\cdot p_4}{\e}\, ,
\end{equation}
is finite if $p_1\cdot p_4\sim\e$. It is most clearly seen in the centre of momentum frame. In the centre of momentum frame, the Regge limit is where the scattering angle $\theta\to0$. We will start with the analytically continued amplitude and use the mass-shell condition $t=-8-s-u$ to replace $t$,
\begin{equation}\label{eq:four-point-full-amplitude-}
   \mathcal{A}^{(4)}(p_1,p_2,p_3,p_4)\sim\frac{\Gamma(4+s/2+u/2)\Gamma(-u/2)}{\Gamma(s/2)}\frac{1}{s^5u^2}\, .
\end{equation}
Using the Stirling's formula we get the known Regge behaviour of the amplitude,
\begin{equation}
    \mathcal{A}^{(4)}(p_1,p_2,p_3,p_4)\sim\left(\frac{s}{2}\right)^{(u/2+4)}\left(1+\frac{u+8}{s}\right)^{s/2}\Gamma(-u/2)\,.
\end{equation}

\subsection{Relating to Gross-Mende saddle}

The scattering amplitude of $n$ tachyon-like particles in the tensionless worldsheet theory is,
\begin{equation}\label{eq:n-point amplitude}
\mathcal{A}^{(n)}_s(p_1,\ldots,p_n)=\frac{(2\pi)^D\tilde{g}^\frac{n}{2}}{\text{Vol~(ISO(2,1))}}\int\prod_{i=1}^ndx_i\braket{\mathcal{V}_{p_1}(x_1,t_1)\ldots\mathcal{V}_{p_n}(x_n,t_n)}\, .
\end{equation}
We can use the global subalgebra $\text{ISO}(2,1)$ of the residual gauge symmetry to fix any three points of insertions. We choose to fix the following three points:
\begin{equation}
    (t_1,x_1)=(0,\infty)\,,~~~(t_2,x_2)=(0,1)\,,~~~(t_n,x_n)=(0,0)\,.
\end{equation}
We then have:
\begin{align}
\mathcal{A}^{(n)}_s(p_1,\ldots,p_n)&=(2\pi)^D\tilde{g}^\frac{n}{2}\int\prod_{i=3}^{n-1}dx_i\braket{\mathcal{V}_{p_1}(t_1,x_1)\ldots\mathcal{V}_{p_n}(t_n,x_n)}\nn\\
&=(2\pi)^D\tilde{g}^\frac{n}{2}\int\prod_{i=3}^{n-1}dx_i\,\exp\left[\frac{c'}{2\e}\mathop{\sum\limits_{i=1}^n\sum\limits_{j=1}^n}_{j\neq i}\,p_i\cdot p_j\,\log{|\mathbf{x}_{ij}|}\right]\,\delta^{(D)}\left(\sum\limits_{i=1}^mp_i\right)\,. 
\end{align}
Since $c'$ is finite and $\{p_i\}$ are finite physical momenta, the exponent in the above integrand is large in the $\e\to0$ limit. The integral hence can be very well-estimated using the saddle point approximation. The exponent $F_n$, which is a function of $(n-3)$ coordinates, 
\begin{equation}
    F_n(x_3,\ldots,x_{n-1})=\frac{c'}{2}\mathop{\sum\limits_{i=1}^n\sum\limits_{j=1}^n}_{j\neq i}\,p_i\cdot p_j\,\log{|x_{ij}|}\, ,
\end{equation}
has the following saddle-point equations,
\begin{equation}
    \nabla_{x_i} F_n(x_3,\ldots,x_{n-1})=0\hspace{2.5mm}\Longrightarrow\hspace{2.5mm}\sum\limits_{\substack{j=1\\j\neq i}}^n\frac{p_i\cdot p_j}{x_i-x_j}=0\, , \quad 2< i< n\,.\label{Scatteringeqn}
\end{equation}
The solutions of the above $(n-3)$ equations will express the $(n-3)$ independent cross-ratios $x_{2<i<n}$ on the worldsheet in terms of the on-shell scattering data on the target space. The above relations \eqref{Scatteringeqn} are known as the scattering equations \cite{Cachazo:2013gna}. For $n=4$, we have the following single scattering equation for the 2-2 scattering,
\begin{align}
   \sum\limits_{{j=1,2,4}}\frac{p_3\cdot p_j}{x_{3}-x_{j}}=0\hspace{2.5mm}\Longrightarrow\hspace{2.5mm}x_3=-\frac{s}{u}\,.\label{4tachyonsaddle}
\end{align}
We find that $x_3=-\frac{s}{u}$ is indeed the point of maximum of $F_4(x_3)$ when $s>0$ and $t,u<0$. 

\medskip

Thus, the saddle point approximation around \eqref{4tachyonsaddle} is simply \eqref{eq:four-point amplitude2o} where we can ignore the constants as compared to the Mandelstam variables,
\begin{align}
\mathcal{A}^{(4)}_s(p_1,\ldots,p_4)&=(2\pi)^D\tilde{g}^2\int\limits_0 ^1dx_3\,\exp\left[\frac{c'}{2\e}\mathop{\sum\limits_{i=1}^4\sum\limits_{j=1}^4}_{j\neq i}\,p_i\cdot p_j\,\log{|x_{ij}|}\right]\,\delta^{(D)}\left(\sum\limits_{i=1}^4 p_i\right)\nn\,,\\
&\sim(2\pi)^D\tilde{g}^2\int\limits_0 ^1dx_3\,\exp{\left[\frac{c'}{\e}\left(p_2\cdot p_3\log{(1-x_3)}+p_3\cdot p_4\log{x_3}\right)\right]}\,\delta^{(D)}\left(\sum\limits_{i=1}^4p_i\right)\nn\,,\\
&\sim(2\pi)^D\tilde{g}^2\exp\left(-\frac{s}{2}\ln s-\frac{t}{2}\ln t-\frac{u}{2}\ln u\right)\delta^{(D)}\left(\sum\limits_{i=1}^4p_i\right)\,.
\end{align}
The above amplitude describes both the leading behaviour of the 2-2 tachyon scattering amplitude in tensionless string theory and the leading Gross-Mende behaviour of usual tensile open string theory amplitudes.

\section{Beyond tensile string theory vertex operators}\label{sec: general-V_p,zeta}

The tensionless string theory amplitudes computed in the previous two sections reflect the universality of the high energy behaviour of string amplitudes. The \textit{tree-level} tensionless amplitudes are simply the correlation function of integrated vertex operators which are obtained from a smooth limit of string theory. However, tensionless string theory can have more classes of vertex operators which are not accessible from a smooth limit of tensile string theory. 

In this section, we consider an extension beyond the usual vertex operators of tensile string theory. We introduce a second class of vertex operators, intrinsically defined on the cylinder, inspired by the extrinsic curvature action of tensionless strings \cite{Savvidy:2004bb},
\begin{align}
    \mathcal{V}_{p_+p_-}=\,:e^{ip_+\cdot X+ip_-\cdot Y}:\, ,\label{newvertex}
\end{align}
where $Y^\mu(\sigma)$ is a time-independent solution of the 2d Carrollian conformal scalar field equation and is related to $X^\mu(\tau,\sigma)$ as,
\begin{align}
    \partial_\sigma Y^\mu(\sigma)=\partial_\tau X^\mu(\tau,\sigma)\, .
\end{align} 

\subsection{The Y-field}
From \eqref{eq: X-mode-expansion}, we obtain the mode-expansion of the field $Y(\sigma)$:
\begin{equation}\label{eq: Y-mode-expansion}
    Y^\mu(\sigma)=y^\mu+\sqrt{\frac{c'}{2}}B^\mu_0\sigma+i\sqrt{\frac{c'}{2}}\sum_{n\neq0}\frac{1}{n}B^\mu_ne^{-in\sigma}\,,
\end{equation}
where the zero-mode $y_0$ is a new mode invisible to the scalar field $X$. The field $Y$ transforms under BMS group similar to the transformation of the scalar $X$ \eqref{eq: X-BMS-transform}. Explicitly, we have,
\begin{align}
    [L_n,Y(\sigma)]=-ie^{in\sigma}\d_\sigma Y\, ,\quad [M_n,Y(\s)]=0\, .
\end{align}
This implies that the plane-mode expansion is,
\begin{equation}
    Y^\mu(x)=y^\mu-i\sqrt{\frac{c'}{2}}B_0^\mu\log(x)+i\sqrt{\frac{c'}{2}}\sum_{n\neq0}\frac{1}{n}B^\mu_nx^{-n}\,.
\end{equation}

In the above we have used another non-trivial commutation relation,
\begin{equation}
    \left[y^\mu\,,\,A^\nu_0\right]=2i\sqrt{\frac{c'}{2}}\eta^{\mu\nu}\,,
\end{equation}
apart from the commutators in \eqref{eq: A-B-commutator}. Interestingly, the two-point function is,
\begin{equation}
    \braket{Y^\mu(\sigma_1)Y^\nu(\sigma_2)}= \braket{y^\mu y^\nu}\, ,
\end{equation}
 just a constant.\footnote{If we ignore the two-point function of the zero modes, similar to the $\braket{XX}$ two-point function, then the two-point function $\braket{YY}$ vanishes.} The scalar nature of the field $Y$ implies that the corresponding vertex operator (with $p^+=0$) $e^{ip^-\cdot Y}$ are the local primaries with $\Delta=\xi=0$.\footnote{We show this explicitly below, see \eqref{eq: new-vertex-2pt-cylinder}.}
 \medskip

Despite the definition of the $Y$-field which is intrinsic to the 2d conformal Carroll scalar, the $Y$-field has its origins in the CFT$_2$. $Y$ is actually the UR limit of the field $X_R-X_L$,
\begin{equation}
    Y(\sigma)=\e \left( X_R-X_L \right)\, ,
\end{equation}
where $X_R$ and $X_L$ are the right moving and left moving parts of the field relativistic scalar $X$,
\begin{equation}\label{XLXR mode}
    \begin{split}
    X^\mu_R(\tau+\sigma)&=x^\mu_R+\sqrt{\frac{\alpha'}{2}}\,\alpha^\mu_{0}(\tau+\sigma)+i\sqrt{\frac{\alpha'}{2}}\sum\limits_{n\in\mathbb{Z}}\frac{1}{n}\,\alpha^\mu_n\,e^{-in(\tau+\sigma)}\, ,\\
    X^\mu_L(\tau-\sigma)&=x^\mu_L+\sqrt{\frac{\alpha'}{2}}\,\tilde\alpha^\mu_{0}(\tau-\sigma)+i\sqrt{\frac{\alpha'}{2}}\sum\limits_{n\in\mathbb{Z}}\frac{1}{n}\,\tilde\alpha^\mu_n\,e^{-in(\tau-\sigma)}\, .
\end{split}
\end{equation}
From the UR limit, we identify the mode expansion as,
\begin{equation}
    Y^\mu(x)=\epsilon(x^\mu_R-x^\mu_L) +\sqrt{\frac{c'}{2}}B_0\sigma+i\sqrt{\frac{c'}{2}}\sum_{n\in\mathbb{Z}}\frac{1}{n}B_ne^{-in\sigma}\, ,
\end{equation}
which implies, $y^\mu=\epsilon(x^\mu_R-x^\mu_L$). The vertex operator in \eqref{newvertex} can thus be understood as the UR limit of the CFT$_2$ vertex operator without a level matching condition. That is the vertex operator,
\begin{equation}
    \mathcal{V}_{p_R,p_L}(\tau,\sigma)=:e^{ip_L\cdot X_L+ip_R\cdot X_R}:\, ,
\end{equation}
with $p_R\neq p_L$ in the UR limit is,
\begin{equation}
    \mathcal{V}_{p^+p^-}=\lim_{\e\to0}\mathcal{V}_{p_R,p_L}\, ,
\end{equation}
where,
\begin{equation}
    p^+=\frac{1}{2}(p_R+p_L)\, ,\quad p^-=\frac{1}{2\e}(p_R-p_L)\,.
\end{equation}
Note that in the induced representation of the tensionless string theory we do not find an intrinsic level matching condition \cite{Bagchi:2021rfw}. The absence of the level matching condition naturally allows the choice $p_R\neq p_L$.

The BMS commutators of the vertex operators $\mathcal{V}_{p^+p^-}$ can be calculated by utilising the $\alpha$-mode expansion,\footnote{The vertex operator with $p_-=0$ gives the original vertex operator studied in the previous sections, while the vertex operator with $p_+=0$ gives a scalar primary.}
\begin{align}
    [M_n\,,\mathcal{V}_{p_+p_-}(\sigma,\tau)]&=-ie^{in\sigma}\partial_\tau \mathcal{V}_{p_+p_-}(\sigma,\tau)\label{MnewVCommC}\, ,\\
    [L_n\,,\mathcal{V}_{p_+p_-}(\sigma,\tau)]&=e^{in\sigma}\left(-i\d_\sigma+n \tau\d_\tau+(\sqrt \e c'\p_+\cdot p_- +\frac{c'\p_+^2}{2}n) \right)\mathcal{V}_{p^+,p^-}(\sigma,\tau)\label{LnewVCommC}\,,
\end{align}
where we need to neglect terms of $\mathcal{O}(p_+^2)\sim \e$ and used the rescaled momenta $\p^+=p^+/\sqrt{\e}$. The term $p_+\cdot p_-$ in the commutator shows that for generic values of $p_-$, the vertex operator is not a primary.\footnote{For finite values of $p^-$, the term $p_+\cdot p_-$ is subleading and can be neglected. The leading order commutators thus reduce to the commutators of the original vertex operators.} To define a primary operator, we demand that $p_+\cdot p_-\sim \e$, that is $p_-$ is orthogonal to $p_+$ upto leading order.

\subsection{Correlators of the new vertex operators}

Once we impose $p_+\cdot p_-=0$, we can proceed to compute the two-point function of the primary vertex operator $\mathcal{V}_{p^+ p^-}$. Similar to the analysis for the usual vertex operators, the product of two vertex operators is computed by using BCH relations between the creation and annihilation modes. The two-point function is simply,
\begin{equation}\label{eq: new-vertex-2pt-cylinder}
\braket{\mathcal{V}_{p^+_1p^-_1}(\tau_1,\sigma_1)\mathcal{V}_{p^+_2p^-_2}(\tau_2,\sigma_2)} = \left(2-2\cos{\sigma_{12}}\right)^{\frac{c'}{2}\p^+_1\cdot \p^+_2}\delta_{\p^+_1+\p^+_2}\delta_{p^-_1+p^-_2}\, ,
\end{equation}
where we have used momentum conservation and ignored the terms of the order $e^{p_+^2}$. The momentum conservation $\delta_{p^-_1+p^-_2}$ can again be understood due to the translational symmetry of the fields $Y$ similar to the momentum conservation $\delta_{p^+_1+p^+_2}$ which occurs due to the translation symmetry of $X$. Since the vertex operator \eqref{newvertex} transforms as a primary, \begin{equation}
 \mathcal{V}_{p_+p_-}(\sigma,\tau)=x^\Delta e^{i\tau\xi}\mathcal{V}_{p_+p_-}(x,t)\, ,
\end{equation}
with $\Delta=c'\left(\p^+\right)^2/2$ and $
\xi=0$, the two-point function on the plane takes the form,
\begin{equation}
    \braket{\mathcal{V}_{p^+_1p^-_1}(t_1,x_1)\mathcal{V}_{p^+_2p^-_2}(t_2,x_2)} = (-x_{12}^2)^{\frac{c'}{2}\p^+_1\cdot \p^+_2}\delta_{\p^+_1+\p^+_2}\delta_{p^-_1+p^-_2}\, .
\end{equation}
where we have the equal weights,
\begin{equation}
\Delta_1=\Delta_2=\frac{c'\left(p^+\right)^2}{2\e}\, ,\quad
\xi_1=\xi_2=0\, .
\end{equation}
with $p^+_1=-p^+_2=p^+$ and $p^-_1=-p^-_2=p^-$. It is then straightforward to show that the higher point functions are,
\begin{equation}
    \braket{\mathcal{V}_{p^+_1p^-_1}(1)\mathcal{V}_{p^+_2p^-_2}(2)\cdots\mathcal{V}_{p^+_1p^-_1}(m)}=\prod_{j>i=1}^n(x_{ij}^2)^{c'\p^+_i\p^+_j/2}\delta\left(\sum_{i=1}^n\p^+_i\right)\delta\left(\sum_{i=1}^n p^-_i\right)\,.
\end{equation}
To construct scattering amplitudes, we require that the integrated vertex operator,
\begin{equation}
    V_{p^+p^-}(\tau)=\int d\sigma \mathcal{V}_{p^+p^-}(\sigma,\tau)\, ,
\end{equation}
is worldsheet diffeomorphism invariant. It is again straightforward to verify that the on-shell condition implies the same conditions as for the usual vertex operators,\footnote{Note that if we regard \eqref{LnewVCommC} and \eqref{MnewVCommC} as basic commutation relations without imposing the orthogonality condition, $p^+\cdot p^-=0$, we recover the orthogonality condition on-shell.}
\begin{equation}
    \Delta=1\, , \quad \xi=0\, .
\end{equation}
The on-shell conditions reproduce,
\begin{equation}
    \left(p^+\right)^2=\frac{2\e}{c'}\, .
\end{equation}
The scattering amplitudes corresponding to these vertex operators thus follow the same arguments as for the vertex operators defined in previous sections. 

\section{Conclusion and discussion}\label{Conclusion and discussion}

\subsection{Summary of results}
The tensionless limit of string theory, as we emphasised in the introduction, is also the very high energy limit of strings. All calculations of amplitudes in the very high energy limit in the literature so far have not been achieved in a true worldsheet computation, where the Carrollian structure of the null worldsheet plays a pivotal role. We have taken a significant step towards understanding this important problem in our work. Let us summarize our main achievements. 

\medskip

We provided the worldsheet description of bosonic tensionless string theory in the induced representation. The tensile string theory is of course built out of the highest weight representation of the residual Virasoro algebra. Quantum null strings constructed in the induced representation are smoothly connected in the tensionless limit of usual (tensile) string theory. Hence this description of null strings is the one we focus on for understanding high energy string scattering. Importantly, the correlators of the highest weight primaries of the closed bosonic string flow to the correlators of the induced representation primaries. We show the match between the limiting analysis and the intrinsic analysis explicitly by using the mode algebra of the free theory.

\medskip

We notice important differences when we compare the vertex operators built on the induced vacuum from existing literature. These induced vertex operators have BMS$_3$ weights different from the vertex operators constructed in the highest weight representation \cite{Hao:2021urq}. The weights in the induced representation $\Delta=c'\p^2/2$ and $\xi=0$ are essentially exchanged from the highest weight representation. We show the computation of the weights explicitly from the mode expansion of the vertex operators. This paves the way to construct, for the first time, worldsheet diffeomorphism invariant scattering amplitudes in tensionless string theory. We computed the 2-2 tachyon scattering amplitude in detail and showed that it reproduced the expected high energy behaviour. 

\medskip

As stressed repeatedly throughout the paper, the high energy limit is equivalent to the large $\alpha'$ limit on the worldsheet. While these arguments have been qualitative and somewhat imprecise in the past, we make this very concrete with an analysis of scattering amplitudes in our paper. Our matching of the tensionless scattering amplitudes, although only tree-level amplitudes, with the high energy amplitudes demonstrated the equivalence at the level of scattering amplitudes. Thus, we re-emphasise
\begin{align}
    \boxed{\text{Scattering of Null strings}_{I} \text{  = String-scattering at very high energies.}} \nonumber 
\end{align}
In the above, Null string$_{I}$ refers to null strings in the induced vacuum. We also emphasise that we have performed, for the first time, an intrinsic analysis on the (null) worldsheet to directly reproduce high energy scattering. 

\subsection{Future directions}

There are a number of directions of research that follow from our analysis in this paper, some of which are extremely important. We list some of these problems to be addressed below. 

\medskip

{$\star$\textit{Tensionless amplitudes beyond tree-level:}} Let us talk about the most immediate question that follows our explorations. This is the pressing problem of the computation of tensionless amplitudes beyond tree-level. To answer that question, we first need to understand the validity of a tensionless string perturbation theory, that is, if we can perform the genus expansion of Carrollian manifolds similar to the genus expansion for Riemann surfaces. Note that the tree-level scattering amplitudes can be explicitly computed in usual, finite tension string theories. By using Stirling's formula, we can expand the string scattering amplitude about the dominant high energy behaviour. The leading behaviour of the tree-level amplitude is the universal Gross-Mende behaviour of the scattering amplitudes common to all string theories. 
We also used Stirling's formula to obtain the tree-level tensionless tachyon string amplitudes. The tensionless string amplitude, at least the tree-level scattering of tachyons, is the same as the Gross-Mende saddle. Since the Gross-Mende saddle is the dominant behaviour of string amplitudes at any order in perturbation theory, the connection between the tensionless tree-level amplitude and the Gross-Mende saddle is important since the tensionless string amplitudes may be treated as universal features as well. A demonstration of the connection beyond tree-level will be very important.  

\medskip

{$\star$\textit{Connections to ambitwistors:}} The null string admits three inequivalent quantizations \cite{Bagchi:2020fpr}, one of which is the null string built on the highest weight representation of the underlying BMS algebra that leads to the ambitwistor strings \cite{Mason:2013sva}. This connection was first noticed in \cite{Casali:2016atr}. Although the vertex operators built in \cite{Hao:2021urq} offers a hint as to how to compute scattering amplitudes from the intrinsic tensionless worldsheet perspective, integrated vertex operators have been a problem \cite{Chen:2025gaz}. We hope to address this with the new class of vertex operators we have introduced in the penultimate section. This is a work in progress. 

\medskip

{$\star$\textit{Amplitudes in the oscillator vacuum:}} The null string has an even more mysterious avatar constructed out of the so-called oscillator vacuum \cite{Bagchi:2020fpr}, which has deep connections to the induced theory \cite{Bagchi:2019cay, Bagchi:2020ats}. The vacua and hence the theories are related by singular Bogoliubov transformations. It would be very interesting to understand what physics hides in the amplitudes of the oscillator null string. 

\medskip

{$\star$\textit{More on the new vertex operators:}} The new vertex operators introduced in Sec~7 hold a lot of promise for explorations beyond conventional string theory into perhaps even the Hagedorn phase. An interesting line of pursuit may be understanding the role of these vertex operators in the context of compactified directions in target space \cite{Banerjee:2024fbi, Banerjee:2023ekd}, where the field $Y$ could play a pivotal role. 

\medskip

{$\star$\textit{Open null strings:}} We have mentioned throughout the paper that the distinction between closed and open null strings blurs in this tensionless limit, and our computations have the flavour of both. This, again as we discussed earlier in the paper, is true for the "null" boundary condition \eqref{nullbc}. There are other open null strings, viz., obeying Dirichlet and Neumann boundary conditions, which appear in the tensionless limit of tensile open strings and have a different underlying symmetry. The vertex operators corresponding to these open string theories would in principle, be very different, and hence it can be expected that the scattering amplitudes would also be different. It would be very interesting to understand these open null string theories and scattering in this context to see how, if at all, these are related to high energy scattering of open strings. 

\medskip

{$\star$\textit{Null superstrings:}} Of course, a natural extension is to null superstrings and analogous problems there. There are two interesting different tensionless limits leading to the homogeneous \cite{Lindstrom:1990qb, Bagchi:2016yyf} and inhomogeneous \cite{Bagchi:2017cte} tensionless superstrings and there would be even more structure to explore here. We should however also interject that a careful canonical quantization akin to the one done for the bosonic string \cite{Bagchi:2020fpr} is also presently lacking for the tensionless superstring. One needs to understand the analogue of the induced vacuum for the superstring and whether it is continuously connected to the usual tensile vacuum first before attempting a scattering amplitude computation.

\subsection*{Acknowledgement}

We thank Aritra Banerjee, Diptarka Das, Dileep Jatkar, Alok Laddha, Shiraz Minwalla, Priyadarshini Pandit, Ashoke Sen, and Stephan Stieberger for important discussions. 

\smallskip

AB is supported partially by a Swarnajayanti Fellowship from Anusandhan National Research Foundation (ANRF) under grant SB/SJF/2019-20/08 and also by an ANRF grant CRG/2022/006165.

\smallskip

SG is partially supported by the grant SB/SJF/2019-20/08, and Institute post-doctoral fellowship from IIT Kanpur. SG thanks Institute of Mathematical Science, IIT and NISER Bhubhaneswar for hospitality where this work was presented as part of the \textit{Quantum fields and Strings} conference and the \textit{Indian Strings Meeting 2025}(ISM 2025) respectively. 

\smallskip

AB, SG, and SRI thank the participants of ISM 2025 for various interesting discussions, especially Shiraz Minwalla and Ashoke Sen for important insights.

\smallskip

AS is supported by the FARE fellowship, IIT Kanpur. SRI is supported by the institute assistantship, IIT Kanpur.

\bigskip \bigskip

\appendix

\section*{APPENDICES}

\section{The Measure}\label{app: measure}
In this section, we will elaborate on the transformation of the measure that arises while defining integrated vertex operators. In the context of textbook string theory, the diffeomorphism invariance of the integrated vertex operator fixes $h=1$ (and $\bar h=1$ for closed strings). In the same light, the invariance of the BMS$_3$ integrated vertex operator under diffeomorphisms should fix the $L_0$ and $M_0$ eigenvalues $\Delta$ and $\xi$, respectively. To this effect, we need to analyse the transformation of the measure under BMS$_3$ transformations obtained as a limit of the Virasoro transformations.

\medskip

Consider the coordinates of CFT$_2$ on the plane $(z,\bar z)$, and let $(\tilde z,\tilde{\bar{z}})$ be the transformed coordinates. In the limiting analysis, we transition from CFT$_2$ on a plane to CCFT$_2$ on a plane via 
\begin{equation}
    z=\epsilon t+x\,,\quad \bar z=\frac{1}{x-\epsilon t}\,.
\end{equation}
Our aim would be to deduce, in a limiting sense, the transformation of the CCFT$_2$ coordinates from the coordinate transformations $z\to z+\varepsilon(z)$ and $\bar z\to\bar z+\bar \varepsilon(\bar z)$. From here on, the $z$ dependence of the function $\varepsilon$ would be assumed implicitly. Therefore, 
\begin{subequations}
    \begin{gather}
        \tilde x=\frac{1}{2}\left(\tilde z+\frac{1}{\tilde{\bar{z}}}\right)\,,\quad\epsilon \tilde t=\frac{1}{2}\left(\tilde z-\frac{1}{\tilde{\bar{z}}}\right)\\
        \tilde x=\frac{1}{2}\left(z+\varepsilon+\frac{1}{\bar{z}+\bar{\varepsilon}}\right)\,,\quad\epsilon \tilde t=\frac{1}{2}\left( z+\varepsilon-\frac{1}{\bar{z}+\bar\varepsilon}\right)\\
        \tilde x=x+\frac{1}{2}\left(\varepsilon-\frac{\bar\varepsilon}{\bar z^2}\right)\,,\quad\epsilon \tilde t=\epsilon t+\frac{1}{2}\left(\varepsilon+\frac{\bar\varepsilon}{\bar z^2}\right)\\
        \tilde x=x+\frac{1}{2}\sum_n\left(\varepsilon_n(\epsilon t+x)^{n+1}-\bar\varepsilon_n(x-\epsilon t)^{-n+1}\right)\,,\nonumber\\\epsilon \tilde t=\epsilon t+\frac{1}{2}\sum_n\left(\varepsilon_n(\epsilon t+x)^{n+1}+\bar\varepsilon_n(x-\epsilon t)^{-n+1}\right)
    \end{gather}
\end{subequations}
In the mode expansion of the anti-holomorphic transformation, replace  $n\to-n$  then take the limit $\epsilon\to0$.
\begin{subequations}
    \begin{gather}
       \tilde x=x+\frac{1}{2}\sum_n(\varepsilon_n-\bar\varepsilon_{-n})x^{n+1}\,,\nonumber\\\epsilon\tilde t=\epsilon t+\frac{1}{2}\epsilon t\sum_n(n+1)(\varepsilon_n-\varepsilon_{-n})x^n+\frac{1}{2}\sum_n(\varepsilon_n+\varepsilon_{-n})x^{n+1}\,,\\
       \tilde x=x+\varepsilon_f(x)\,,\quad \epsilon\tilde t=\epsilon t\varepsilon_{f}'(x)+\varepsilon_g\,,\\
       \text{where,}~~~\varepsilon_f=\frac{1}{2}\sum_n(\varepsilon_n-\bar\varepsilon_{-n})x^{n+1}~~~\text{and}~~~\varepsilon_g=\frac{1}{2}\sum_n(\varepsilon_n+\bar\varepsilon_{-n})x^{n+1}\,,\\
       \text{therefore,}~~~\tilde x=f(x)~~~\text{and}~~~\tilde t=tf'(x)+g(x)\,.
       \end{gather}
\end{subequations}
To derive the transformation of the measure, lets first note down the Cauchy-Riemann condition with $t$ replaced with $\epsilon t$, without taking the limit on $\epsilon$. The Cauchy Riemann conditions are given as $$\frac{\partial\tilde t}{\partial t}=\frac{\partial \tilde x}{\partial x}\,,\quad \epsilon\frac{\partial\tilde t}{\partial x}=-\frac{\partial \tilde x}{\epsilon\partial t}\,.$$
The measure part of the integrated vertex operator transforms as
\begin{subequations}
    \begin{gather}
        \epsilon\int d\tilde td\tilde x=\int
        \begin{pmatrix}
         \frac{\partial\tilde t}{\partial t}&\epsilon\frac{\partial\tilde t}{\partial x}\\\frac{\partial\tilde x}{\epsilon\partial t}&\frac{\partial\tilde x}{\partial x}   
        \end{pmatrix}\epsilon dtdx\\
        \text{Using the Cauchy Riemann conditions, we get,}~~
        \epsilon\int d\tilde td\tilde x=\int
        \begin{pmatrix}
         \frac{\partial \tilde x}{\partial x}&\epsilon\frac{\partial\tilde t}{\partial x}\\-\epsilon\frac{\partial\tilde t}{\partial x}&\frac{\partial\tilde x}{\partial x}   
        \end{pmatrix}\epsilon dtdx\\
        \text{Using the transformations derived above and taking the $\epsilon\to0$ limit, we get,}\nonumber\\
        \epsilon\int d\tilde td\tilde x=\epsilon \int dtdx(1+2\varepsilon_f'(x))=\epsilon \int dtdxf'(x)^2\,.
    \end{gather}
\end{subequations}
Interestingly, we do not get any dependence on $g(x)$, which, generates the $M_n$ charges in the BMS$_3$ algebra. Therefore, one should expect that the measure does not contribute to any particular condition on $\xi$. This implies that the factors of $\xi$, which will come in the transformed integrated vertex operators, have to be set to zero, or they should cancel among themselves. It can also be shown explicitly that the measure is boost ($M_0$) invariant.

\section{AB normal ordering}\label{app: AB-normal-order}
In the induced representation the $B$-modes annihilate the induced vacuum,
\begin{equation}
    B_n\ket{0}=0\, ,\quad\forall n\in \mathbb{Z}\, .
\end{equation}
The above property would prompt to define the $AB$ normal ordering, 
\begin{equation}
    :A_nB_m:=A_nB_m\, ,\quad \forall n,m\in \mathbb{Z}\, ,
\end{equation}
where the $B$-modes are kept to the right. However note that the $B$-modes also annihilate the left vacuum,
\begin{equation}
    \bra{0}B_n=0\, ,\quad\forall n\in \mathbb{Z}\, .
\end{equation}
The $B$-modes treat the left and right vacuua on an equal footing. This poses a problem in defining a normal ordering using the $AB$ modes.
If we continue with the $AB$ normal ordering, we have the following definition of the vertex operators,
\begin{equation}
\mathcal{V}_p[\Phi(\sigma,\tau)]= :\exp\left(ip\,\Phi(\sigma,\tau)\right):=\exp\left(ip\left(\phi+i\sum_{n\neq0}\frac{1}{n}A_ne^{-in\sigma}\right)\right)\exp\left(ip\tau\sum_{n\in\mathbb{Z}}B_me^{-in\sigma}\right)\, ,
\end{equation}
where we have understood the creation and annihilation modes as,
\begin{align}
e^{C}&=\exp\left(ip\left(\phi+i\sum_{n\neq0}\frac{1}{n}A_ne^{-in\sigma}\right)\right)\\
e^{A}&=\exp\left(ip\tau\sum_{n\in\mathbb{Z}}B_n e^{-in\sigma}\right)
\end{align}
The product of two vertex operators, 
\begin{align}
    \mathcal{V}_{p_1}\mathcal{V}_{p_2}&=:\exp\left(ip\,\Phi(1)\right)::\exp\left(ip\,\Phi(2)\right):\, ,\nonumber\\
    &=e^{C(1)}e^{C(2)}e^{A(1)}e^{A(2)}\exp\left(-p_1p_2\tau_1\left[\sum_{n\in\mathbb{Z}}B_ne^{-in\sigma_1},\phi+i\sum_{m\neq0}\frac{1}{m}A_me^{-im\sigma_2}\right]\right)\, ,\nonumber\\
    &=\mathcal{V}_{p_1+p_2}\exp\left(i\frac{c'}{2}p_1p_2\tau_1\delta(\sigma_1-\sigma_2)\right)\, .
\end{align}
Notice that the two-point function,
\begin{equation}
     \mathcal{V}_{p_1}\mathcal{V}_{p_2}=\delta_{p_1+p_2}\exp\left(i\frac{c'}{2}p_1p_2\tau_1\delta(\sigma_1-\sigma_2)\right)\, ,
\end{equation}
is not translational invariant on the cylinder. It depends rather asymmetrically on the time coordinate of the first vertex operator. Probably that is the least of the worries the above correlator presents. 

\subsection*{\texorpdfstring{BMS$_3$}{BMS3} weights of the vertex operator in AB-normal ordering}

However, in the AB-normal ordering, i.e with the $B$ modes to the right, we have,
\begin{align}
    [L_n,\mathcal{V}_p]&=\frac{1}{2}\sum_{m\in\mathbb{Z}}\left([A_{-m},\mathcal{V}_p]B_{m+n}+A_{-m}[B_{m+n},\mathcal{V}_p]\right)\nn\\
    &=p\sqrt{\frac{c'}{2}}\sum_{m\in\mathbb{Z}}\left(-im\tau\mathcal{V}_pB_{m+n}e^{-im\sigma}+A_{-m}\mathcal{V}_pe^{i(m+n)\sigma)}\right)\, .
\end{align}
Notice that the above expression on the right hand side is already normal ordered. Thus, the commutator is just,
\begin{equation}
    [L_n,\mathcal{V}_p]=e^{in\sigma}\left(-i\partial_\sigma+n\tau \d_\t\right)\mathcal{V}_p\, .
\end{equation}
The above commutator suggests that $\Delta=0$ and $\xi=0$. However we have,
\begin{equation}
    [M_n,\mathcal{V}_p(x,t)]=e^{in\sigma}\left(-i\d_\t+\infty\,\frac{p^2c'}{2}\right)\mathcal{V}_p(x,t)\,,
\end{equation}
demonstrating that $\xi$ of $\mathcal{V}_p$ is not well-defined wrt the AB normal ordering. The general vertex operator $\mathcal{V}_{p^+p^-}$ suffers from the same problem.

\section{BMS weights of the vertex operator}\label{app: BMS-weights}
In the previous section, we saw that AB normal ordering leads to an ill-defined $\xi$ for the BMS$_3$ vertex operator. In this appendix, we rectify this issue. To compute the BMS$_3$ eigenvalues of the vertex operators we need to compute the commutators of the vertex operator with $L_0$ and $M_0$. We will calculate the commutators of the vertex operator on the cylinder. 

\medskip

The computation of the commutator relies on a regularisation scheme which is different from the (generalized) $\zeta$-function regularisation, since the latter is not very suitable for our purpose. As an example, we will first show this process for the vertex operator in the usual 2d CFT and then for our case of 2d CCFT.

\subsection{CFT regulators and commutator calculations}

We motivate the use of a regulator for computing commutators in 2d CCFT from the computation of the commutator in 2d CFT. In this subsection, the vertex operator in 2d CFT will be denoted as $\mathcal{V}^{\text{CFT}_2}_p(z,\bar z)\equiv\mathcal{V}_p$. The commutator of the vertex operator with the holomorphic Virasoro generators $\mathcal{L}_n$ is given as 
\begin{equation}\label{eq:CFT calc step1}
    [\mathcal{L}_n,\mathcal{V}_p]=\frac{1}{2}\sum_{m\in\mathbb{Z}}[\alpha_{-m}\alpha_{m+n},\mathcal{V}_p]=\frac{1}{2}\sum_{m\in\mathbb{Z}}\left([\alpha_m,\mathcal{V}_p]\alpha_{-m+n}+\alpha_{-m+n}[\alpha_m\mathcal{V}_p]\right)\,.
\end{equation}
Using the commutators
\begin{equation}
    [\alpha_n,\mathcal{V}_p]=\sqrt{\frac{\alpha'}{2}}pz^n \mathcal{V}_p\, ,\quad [\tilde \alpha_n,\mathcal{V}_p]=\sqrt{\frac{\alpha'}{2}}p\bar z^n\mathcal{V}_p\, ,
\end{equation}
in \eqref{eq:CFT calc step1} we get
\begin{align}\label{eq: [L,V]inCFT-no-reg-step}
    [\mathcal{L}_n,\mathcal{V}_p]&=\sqrt{\frac{\alpha'}{2}}\frac{1}{2}\sum_{m}\left(pz^{m}\mathcal{V}_p\alpha_{n-m}+\alpha_{n-m}pz^{m}\mathcal{V}_p\right)
    \nn\\&=\sqrt{\frac{\alpha'}{2}}\frac{p}{2}\sum_{l}\left(\mathcal{V}_p\alpha_l+\alpha_l\mathcal{V}_p\right)z^{n+1}z^{-l-1}\, ,
\end{align}
focusing only on the part which requires normal ordering,
\begin{align}
    [\mathcal{L}_n,\mathcal{V}_p]&\sim \sqrt{\frac{\alpha'}{2}}\frac{p}{2}z^{n}\left(\sum_{l\geq 0}[\alpha_l,\mathcal{V}_p]z^{-l}-\sum_{l<0}[\alpha_l,\mathcal{V}_p]z^{-l}\right)\nn\,,
\end{align}
we chose the regulator as follows. We begin at the step,
\begin{equation}
    [\mathcal{L}_n,\mathcal{V}_p]\sim\sqrt{\frac{\alpha'}{2}}p\left(z^{n+1}\sum_{l<0}[\mathcal{V}_p,\alpha_l]z^{-l-1}+z^{n+1}\sum_{l\geq 0} z^{-l-1}[\alpha_l,\mathcal{V}_p]\right)\, .
\end{equation}
We will regulate the summations as follows
\begin{align}
    [\mathcal{L}_n,\mathcal{V}_p]\sim{\frac{\alpha'p^2}{4}}\left(-z_1^{n+1}\sum_{l<0}z_1^{-l-1} z_2^{l}+z_2^{n+1}\sum_{l\geq 0}z_1^{l} z_2^{-l-1}\right)\mathcal{V}_p=\frac{\alpha' p^2}{4z_{12}}\left(z_1^{n+1}-z_2^{n+1}\right)\mathcal{V}_p\,.
\end{align}
Now, taking the limit $z_1\to z_2=z$, we have,
\begin{equation}
    [\mathcal{L}_n,\mathcal{V}_p]=z^{n+1}\partial_z \mathcal{V}_{p}+\frac{p^2\alpha'}{4}(n+1)z^n\mathcal{V}_p\,~~\forall n\in\mathbb{Z},
\end{equation}
here we have re-inserted the vector action. 

\subsection{BMS regulators and commutator calculation}
In this section, we will derive the transformation of the BMS vertex operator on the cylinder\footnote{Note that we have used the same notation $\mathcal{V}_p$ for the vertex operator, but in this section it is used in the context of BMS. Moreover, the vertex operator in this section is defined on the cylinder.} $\mathcal{V}_p^{\text{BMS}_3}(\s,\t)\equiv\mathcal{V}_p$. Inspired by the previous section, we will show that a similar regularisation scheme can be employed to derive the commutator brackets.

\subsection*{$[L_n,\mathcal{V}_p]$ computation:}
The $L_n$ generators admit the mode expansion given in \eqref{eq:BMS3 gen in modes}. Generators with $n\neq0$ do not need a normal ordering prescription. However, for $L_0$, normal ordering in particular is important. We use the $AB$ normal ordering, i.e., $B$ modes to the right, in the definition of $L_0$, which is equivalent to the $\alpha$ normal ordering (see \cref{sec:The induced representation}). The commutator of the vertex operator with the generators is given as
\begin{equation}\label{eq: Ln Vp commutator}
[L_n,\mathcal{V}_p]=\frac{1}{2}\sum_{m\in\mathbb{Z}}[A_{-m+n}\cdot B_{m},\mathcal{V}_p]\,
\end{equation}
We use the following commutators,
\begin{equation}
[A_m^\mu,\mathcal{V}_p]=im\t p^\mu\sqrt{2c'}  e^{im\sigma}\mathcal{V}_p\, ,\quad[B_m^\mu,\mathcal{V}_p]=p^\mu\sqrt{2c'} e^{im\sigma}\mathcal{V}_p\, ,
\end{equation}
in \cref{eq: Ln Vp commutator} to get
\begin{equation}\label{eq: [l0,v]-A-B-modes}
     [L_n,\mathcal{V}_p]=p\sqrt{\frac{c'}{2}}\sum_{m\in\mathbb{Z}}\left(-im\tau\mathcal{V}_pB_{m+n}e^{-im\sigma}+A_{-m}\mathcal{V}_pe^{i(m+n)\sigma)}\right)\, .
\end{equation}
The right-hand side is not in a normal ordered form when expressed in terms of the $\alpha$- oscillators. 

Using \eqref{eq: AB-to-alphas} in the above expression and following the normal ordering prescription for the $\alpha$-oscillators, we get
\begin{align}\label{eq: Ln Vp 1}
    [L_n,\mathcal{V}_p]&=p\sqrt{\frac{c'}{2}}\sum_{m\in\mathbb{Z}}\left(-im\tau:\mathcal{V}_pB_{m+n}:e^{-im\sigma}+:A_{-m}\mathcal{V}_p:e^{i(m+n)\sigma)}\right)\,\nn\\
    &\qquad+p\sqrt{\frac{c'}{2\epsilon}}\left(-i\tau\epsilon\Bigg(\sum_{m<-n}m\mathcal{V}_p\alpha_{m+n}e^{-im\sigma}+\sum_{m>-n}m\mathcal{V}_p\alpha_{-m-n}e^{-im\sigma}\right)\nn\\
    &\qquad+\sum_{m\leq0}\alpha_{-m}\mathcal{V}_pe^{i(m+n)\sigma}-\sum_{m\geq0}\alpha_{m}\mathcal{V}_pe^{i(m+n)\sigma}\Bigg)\, .
\end{align}
where the first line is the normal ordered part, and the remaining part is to be normal ordered. 

To bring the above expression into a normal ordered form, we need to use the following commutators
\begin{equation}\label{eq: [a,V]-commutator}
    [\alpha_n,\mathcal{V}_p]=p\sqrt{\frac{c'}{2\epsilon}}e^{in\sigma}\, \mathcal{V}_p\left(1+\mathcal{O}(\e)\right)\, ,\quad
    [\tilde\alpha_n,\mathcal{V}_p]=p\sqrt{\frac{c'}{2\epsilon}}e^{-in\sigma}\, \mathcal{V}_p\left(1+\mathcal{O}(\e)\right)\, .
\end{equation}
The above commutators are essentially the UR limit of the corresponding CFT$_2$ commutators. Now, on substituting \eqref{eq: [a,V]-commutator} into \eqref{eq: Ln Vp 1}, keeping only the leading terms and introducing the regulator, \cref{eq: Ln Vp 1} takes the following form,
\begin{multline}
    [L_n,\mathcal{V}_p]=e^{in\sigma}\left(-i\partial_\sigma+n\tau \d_\t\right)\mathcal{V}_p\\+p^2\frac{c'}{2\epsilon}\left(e^{i(n+1)\sigma_2}\sum_{m\geq0}e^{i(-m-1)\sigma_2}e^{im\sigma_1}-e^{i(n+1)\sigma_1}\sum_{m\geq0}e^{-im\sigma_2}e^{i(m-1)\sigma_1}\right)\mathcal{V}_p\,.
\end{multline}
We use the cylinder to plane coordinate transformation\footnote{Note that this coordinate transformation is employed solely to facilitate the summation; the commutator itself is still computed on the cylinder.} \eqref{eq: cylinder-to-plane-coords} only in the last two terms to get,
\begin{equation}
    [L_n,\mathcal{V}_p]=e^{in\sigma}\left(-i\partial_\sigma+n\tau \d_\t\right)\mathcal{V}_p+p^2\frac{c'}{2\epsilon}\left(x_2^{n}\sum_{m\geq0}\left(\frac{x_1}{x_2}\right)^m-x^n_1\sum_{m\geq0}\left(\frac{x_1}{x_2}\right)^m\right)\mathcal{V}_p\,.
\end{equation}

Performing the summation, taking the limit $x_1\to x_2=x$ and converting back to the cylinder coordinates we obtain,
 \begin{equation}\label{eq: LnVp-intrinsic}
 [L_n,\mathcal{V}_p]=\left(-ie^{in\sigma}\partial_\sigma+n\tau e^{in\sigma}\d_\t+ne^{in\sigma}\frac{c'p^2}{2\epsilon}\right)\mathcal{V}_p\, .\,
 \end{equation}
By setting $n=1$ in the above commutator, we can read off $\Delta$ and $\xi$ as follows,
\begin{equation}
	\Delta=\frac{c'p^2}{2\epsilon}=\frac{c'\p^2}{2}\,,\quad \xi=0\,.
\end{equation}
where we have used the rescaled momenta, $\p=p/\sqrt \e$.

\subsection*{$[M_n,\mathcal{V}_p(x,t)]$ computation}
The $M_n$ generators admit the mode expansion given in \eqref{eq:BMS3 gen in modes}. Here we do not need to use any normal ordering even for $M_0$. The commutator of the vertex operator with the $M_n$ generators is given as 
\begin{align}
	[M_n,\mathcal{V}_p]&=\frac{1}{4}\sum_{m\in\mathbb{Z}}[B_{-m}B_{m+n},\mathcal{V}_p]=\frac{1}{4}\sum_{m\in\mathbb{Z}}(B_{-m}\cdot[B_{m+n},\mathcal{V}_p]+[B_{-m},\mathcal{V}_p]\cdot B_{m+n})\, .
\end{align}
At this step we use the mode expansion of the vertex operator in terms of the $A,B$ modes to get,
\begin{equation}
	[M_n,\mathcal{V}_p(x,t)]=\sqrt{\frac{c'}{8}}p\sum_{m\in\mathbb{Z}}\left(\mathcal{V}_pB_{m+n}e^{-im\sigma}+B_{-m}\mathcal{V}_pe^{i(m+n)\sigma}\right)\, .
\end{equation}
Similar to the previous computation, we will express the $B$- modes in terms of the $\alpha$-oscillators \eqref{eq: AB-to-alphas} and normal order the expression with respect to the $\alpha$-oscillators to arrive at
\begin{multline}
[M_n,\mathcal{V}_p]=-ie^{in\sigma}\partial_\t\mathcal{V}_p+p\sqrt{\frac{\epsilon c'}{8}}\Bigg(\sum_{m<-n}\mathcal{V}_p\alpha_{m+n}e^{-im\sigma}+\sum_{m\leq0}\alpha_{-m}\mathcal{V}_pe^{i(m+n)\sigma}\\+\sum_{m>-n}\mathcal{V}_p\tilde\alpha_{-m-n}e^{-im\sigma}+\sum_{m\geq0}\tilde\alpha_{m}e^{i(m+n)\sigma}\mathcal{V}_p\Bigg)\, .
\end{multline}

We use the commutators \eqref{eq: [a,V]-commutator} in the above equation and convert the summations from the cylinder to the plane coordinates. Furthermore, we will also put in our regularisation scheme to obtain,
\begin{align}\label{eq: Mn-Vp-intrinsic-comm}
[M_n,\mathcal{V}_p]&=-ie^{in\sigma}\partial_\t\mathcal{V}_p+\frac{c'p^2}{4}\Bigg(-x_1^{n+1}\sum_{m<0}x_2^mx_1^{-m-1}+x_2^{n+1}\sum_{m\leq0}x_1^{-m}x_2^{m-1}\nn\\&\qquad\qquad\qquad\qquad\qquad\quad-x_2^{n+1}\sum_{m>0}x_1^mx_2^{-m-1}+x_1^{n+1}\sum_{m\geq0}x_2^{-m}x_1^{m-1}\Bigg)\mathcal{V}_p\, ,\nonumber\\
    &=-ie^{in\sigma}\partial_\t\mathcal{V}_p+\e \frac{c'\p^2}{2}e^{in\sigma}\mathcal{V}_p\,.
\end{align}
To get the second equality, we have taken the limit $x_2\to x_1=x$ and substituted $\epsilon\p^2$ in place of $p^2$.

We learn that\footnote{Note that the second term in the last line of \cref{eq: Mn-Vp-intrinsic-comm} is subleading.} $\xi=0$.

\medskip

We have now established via \cref{eq: LnVp-intrinsic,eq: Mn-Vp-intrinsic-comm} that the vertex operator in the induced representation is a BMS primary with weights
\begin{equation}
\Delta=\frac{\p^2 c'}{2}\, ,\quad \xi=0\, .
\end{equation}
Having calculated the BMS weights for the usual vertex operator, we will follow the same procedure to calculate the weights of the new vertex operator \eqref{newvertex}.

\medskip

To compute the commutator of the general vertex operator \eqref{newvertex} with the BMS generators, the following commutators will be useful
\begin{subequations}\label{eq: alpha and newvertex comm}
\begin{align}
   [\alpha_n\,,\mathcal{V}_{p_+p_-}]=&\sqrt{\frac{c'}{2\epsilon}}\left[p_++\epsilon\left(p_-+in\tau p_+\right)\right]e^{in\sigma}\mathcal{V}_{p_+p_-}\,,\\
   [\tilde{\alpha}_{-n}\,,\mathcal{V}_{p_+p_-}]=&\sqrt{\frac{c'}{2\epsilon}}\left[p_+-\epsilon\left(p_-+in\tau p_+\right)\right]e^{in\sigma}\mathcal{V}_{p_+p_-}\,,
\end{align}
\end{subequations}
Using the commutators between the $\alpha$-oscillators and the vertex operator given above, and our regularisation scheme, we get
\begin{align}
    &[L_n,\mathcal{V}_{p_+p_-}]=\left(-ie^{in\sigma}\partial_\sigma+n\tau e^{in\sigma}\partial_\tau+e^{in\sigma}c'p_+\cdot p_-+ne^{in\sigma}\frac{c'p_+^2}{2\epsilon}\right)\mathcal{V}_{p_+p_-}\,,\label{eq: Ln-newvertex}\\
    &[M_n,\mathcal{V}_{p_+p_-}]=-ie^{in\sigma}\d_\t\mathcal{V}_{p_+p_-}+\epsilon\frac{c'\p^2_+}{2}e^{in\sigma}\mathcal{V}_{p_+p_-}\,.\label{eq: Mn-newvertex}
\end{align}
Note that the second term in \eqref{eq: Mn-newvertex} is subleading. Therefore, from \cref{eq: Ln-newvertex,eq: Mn-newvertex}, we conclude that the new vertex operator has the following BMS weights
\begin{equation}
    \Delta=\frac{c'\p_+^2}{2}\,,\quad \xi=0\,.
\end{equation}

\section{Integrated ``closed string'' vertex operator}\label{app: closed-type-integrated-Vp}

In section \ref{sec: scattering} we defined the integrated vertex operators in close analogy with open strings. We can also define the integrated vertex operator as,
\begin{equation}\label{def: int-vertex-op-c}
    V_p=\tilde{g}_s\int d^2\xi\,\mathcal{V}_p(\xi_i) \,,
\end{equation}
where $\tilde g_s$ is a coupling constant. \footnote{We have added subscript $s$ to distinguish from the coupling $\tilde g$ defined with the integrated vertex operator in \eqref{intvert}.}
The diffeomorphism invariance of the integrated vertex operator implies $\mathcal{V}_p(x,t)$ has,
\begin{equation}
    \Delta=2\implies\lim_{\epsilon\to0} p^2=\frac{4\epsilon}{c'}\, ,\quad \xi=0\, .
\end{equation}
In particular, the tachyon state of tensile string theory becomes massless in the tensionless limit. Note that this again demonstrates a smooth limit $\epsilon\to0$. We now present the scattering amplitude analysis of tensionless strings as would have been with the above vertex operator \eqref{def: int-vertex-op-c} and the problems associated with this definition.

\subsection*{Scattering amplitudes}

We show the three-point amplitude for completeness and move to the important tree-level four-point scattering amplitude. 

\subsubsection*{Three-point scattering amplitude:}
The three-point amplitude of the tachyon-like vertex operators is,
\begin{equation}\label{eq:three-point amplitude}
    \mathcal{A}^{(3)}(p_1,p_2,p_3)=\frac{(2\pi)^D\tilde{g}_s^{3/2}}{\text{Vol~(ISO(2,1))}}\int\prod_{i=1}^3d^2\xi_i\braket{\mathcal{V}_{p_1}(\xi_1)\mathcal{V}_{p_2}(\xi_2)\mathcal{V}_{p_3}(\xi_3)}\, .
\end{equation}

The three-point function  of the vertex operators is (see \eqref{eq: three-point-vertexes})
\begin{equation}
    \braket{\mathcal{V}_{p_1}(1)\mathcal{V}_{p_2}(2)\mathcal{V}_{p_3}(3)}=(x_{12}^2)^{\p_1\p_2c'/2}(x_{23}^2)^{\p_2\p_3c'/2}(x_{13}^2)^{\p_1\p_3c'/2}\delta(\p_1+\p_2+\p_3)\,. 
\end{equation}

Using the residual symmetry to fix the three positions, the three-point correlator is given as, 
\begin{align}
    \braket{\mathcal{V}_{p_1}(0,0)\mathcal{V}_{p_2}(0,1)\mathcal{V}_{p_3}(0,\infty)}
    &=(-1)^{-\p_3^2c'}\delta(\p_1+\p_2+\p_3)\,,
\end{align}
The on-shell condition of the physical momentum $\p_3^2 c'=4$ implies,
\begin{equation}
    \braket{\mathcal{V}_{p_1}(0,0)\mathcal{V}_{p_2}(0,1)\mathcal{V}_{p_3}(0,\infty)}=\delta(\p_1+\p_2+\p_3)\, .
\end{equation}
Substituting the three-point correlator along with the positions of the vertex operators in \eqref{eq:three-point amplitude}, we get,
\begin{equation}
        \mathcal{A}^{(3)}(p_1,p_2,p_3)=(2\pi)^D\tilde{g}_s^{3/2}\delta(\p_1+\p_2+\p_3)\,.
\end{equation}
We now proceed to the non-trivial four-point scattering amplitude.

\subsubsection*{Four-point scattering amplitude:}

The 2-2 scattering amplitude of tachyon-like vertex operators is,
\begin{equation}\label{eq:four-point amplitude1}
    \mathcal{A}^{(4)}_s(p_1,p_2,p_3,p_4)=\frac{(2\pi)^D\tilde{g}_s^2}{\text{Vol~(ISO(2,1))}}\int\prod_{i=1}^4d^2\xi_i\braket{\mathcal{V}_{p_1}(x_1,t_1)\dots\mathcal{V}_{p_4}(x_4,t_4)}\, .
\end{equation}
The four-point function of the vertex operators is (see \eqref{eq: n-point-vertexes}),
\begin{equation}\label{eq: four-point correlator 1}
	\braket{\mathcal{V}_{p_1}(1)\cdots\mathcal{V}_{p_4}(4)}=\delta\left(\sum_{i=1}^{4}p_i\right)\prod_{i<j=1}^{4}(x_{ij}^2)^{\p_i\p_j c'/2}\, .
\end{equation}
We use the global subgroup $ISO(2,1)$ to fix the positions of three of the four vertex operators and also order them similar to \eqref{eq: x ordered vertexop}. 
The four point correlator \eqref{eq: four-point correlator 1} is then given as,

\begin{align}\label{eq: four-point correlator 2}
    \braket{\mathcal{V}_{p_1}(1)\cdots\mathcal{V}_{p_4}(4)}&=
    \delta\left(\sum_{i=1}^{4}p_i\right)x^{-4-s/2}(1-x)^{-4-u/2}\,,
\end{align}
where $x$ is the cross-ratio. The above equation represents one of the partial amplitudes while the total amplitude is the sum of the $s\, ,t\, ,u$-channels. The Mandelstam variables have been defined in \cref{Mandelstam variables} and the mass-shell condition is given as 
\begin{equation}\label{eq: closed strings mass shell condition}
    s+t+u=-16\,.
\end{equation}

The Mandelstam variables are large as they scale as $1/\e$ for finite $p_i$. However, like before, we will continue with the computation without using the large momenta asymptotics and take the limit at the end of the computation. 

The four-point amplitude in the integral form is found to be
\begin{equation}\label{eq:four-point amplitude2}
\mathcal{A}^{(4)}(p_1,p_2,p_3,p_4)=(2\pi)^D\tilde{g}_s^2\delta\left(\sum_{i=1}^{4}p_i\right)\underbrace{\int_0^1dx~x^{-4-s/2}(1-x)^{-4-u/2}}_{I(s,u)}\,.
\end{equation}
The integral converges in the region $\Re(s)<-6$ and $\Re(u)<-6$. As $s\to-6$, the divergence can be analysed by approximating the integrand $I(s,u)$ in a neighbourhood of $x=0$
\begin{equation}
I(s,u)=-\frac{r^{-3-s/2}}{s/2-3}~+\text{terms analytic at}~s=-6\,,
\end{equation}
where $r$ is the upper limit of the integral $I(s,u)$. Therefore, the s-channel pole is located at $s=-6$. The integral \eqref{eq:four-point amplitude2} evaluates to,
\begin{equation}
\mathcal{A}^{(4)}(p_1,p_2,p_3,p_4)=(2\pi)^D\tilde{g}_s^2\delta\left(\sum_{i=1}^{4}p_i\right)\frac{\Gamma(-3-s/2)\Gamma(-3-u/2)}{\Gamma(2+t/2)}\,,
\end{equation}
with $\Re(s)\,,~$$\Re(u)<-6$. Adding all the partial amplitudes, we obtain the full crossing-symmetric amplitude as follows
\begin{multline}\label{eq:four-point amplitude3}
    \mathcal{A}^{(4)}(p_1,p_2,p_3,p_4)=(2\pi)^D\tilde{g}_s^2\delta\left(\sum_{i=1}^{4}p_i\right)\Bigg(\frac{\Gamma(-3-s/2)\Gamma(-3-u/2)}{\Gamma(2+t/2)}\\
    +\frac{\Gamma(-3-s/2)\Gamma(-3-t/2)}{\Gamma(2+u/2)}+\frac{\Gamma(-3-t/2)\Gamma(-3-u/2)}{\Gamma(2+s/2)}\Bigg)\,.
\end{multline}

The above scattering amplitude again looks strikingly similar to the Veneziano amplitude in open tensile string theory with large Mandelstam variables. 

The physical region of scattering corresponds to $s\to\infty$, $t,u\to-\infty$. But the region of convergence of the integral \eqref{eq:four-point amplitude2} is different from the physical region. Therefore, analytically continuing \eqref{eq:four-point amplitude3} to the physical region, $\Re(s)>0$ and $\Re(t)$, $\Re(u)<-6$, we get,

\begin{multline}\label{eq:four-point-full-amplitude-closed}
   \mathcal{A}^{(4)}(p_1,p_2,p_3,p_4)=(2\pi)^D\tilde{g}_s^2\delta\left(\sum_{i=1}^{4}p_i\right)\frac{\Gamma(-t/2)\Gamma(-u/2)}{\Gamma(s/2)}\frac{256}{(2+s)(2+t)(2+u)s}\times\\
   \times\left[\frac{\sin(\pi t/2)}{\sin(\pi s/2)}\frac{1}{(6+s)(4+s)(6+u)(4+u)}+\frac{\sin(\pi u/2)}{\sin(\pi s/2)}\frac{1}{(6+s)(4+s)(6+t)(4+t)}\right.\\
    \left.+\frac{1}{(6+t)(4+t)(6+u)(4+u)}\right]  
\end{multline}
The pole structure is now clearly separated. The poles are located at the values $s=2n$, with $n\in\mathbb{Z}_{\geq -3}$. The first pole has a state with $m'^2=-6/c'$. This state is however not there in the spectrum. The interpretation of this pole is still unclear. It could also be a fake pole. If it corresponds to a state, then it is clearly a lighter state than the tachyon-like state itself. The first excited state produces the correct mass,
\begin{equation}\label{eq: closed-tachyon-mass}
    m'^2=-\frac{4}{c'}
\end{equation}
The remaining poles correspond to the higher excited states,
\begin{equation}\label{eq: closed-mass-spectrum}
    m'^2=-\frac{2}{c'}(2-n)\,\quad n\in\mathbb{Z}_{> 0} .
\end{equation}
We will now use largeness of the Mandelstam variables to obtain the leading behaviour by using Stirling's formula, \eqref{eq:Stirling's Formula}.
The leading behaviour is found to be
\begin{multline}
    \frac{\Gamma(-t/2)\Gamma(-u/2)}{\Gamma(s/2)}=
    -2\pi^{1/2}e^{-8}(-1)^{s/2} \left(\frac{s}{tu}\right)^{1/2}t^{-t/2} u^{-u/2} s^{-s/2}\\
    \times\exp\left[\sum_{k=1}^\infty\frac{2^{2k-1}\zeta(1-2k)}{1-2k}\left(t^{1-2k}+u^{1-2k}+s^{1-2k}\right)\right]\,,
\end{multline}
where we have used the mass-shell condition \eqref{eq: closed strings mass shell condition}.

\medskip

At this stage, the analysis still allows for finite-angle as well as infinitesimal scattering angle $\theta$ in the centre of mass frame. In other words, we can obtain both the Gross-Mende and the Regge \cite{Gross:1987kza} behaviour of the amplitude \eqref{eq:four-point amplitude3}. \\

\emph{Hard scattering limit}: 
Using the largeness of the Mandelstam variables, we find the asymptotic form of the amplitude to be,
\begin{multline}\label{eq: hard-scattering}
    \mathcal{A}^{(4)}(\{p_i\})\sim(-1)^{\frac{s}{2}-1}e^{-\frac{s}{2}\ln(s/2)-\frac{t}{2}\ln(t/2)-\frac{u}{2}\ln(u/2)}\left(stu\right)^{-3/2}\\\times\left[\frac{\sin(\pi t/2)}{\sin(\pi s/2)}\frac{1}{s^2u^2}+\frac{\sin(\pi u/2)}{\sin(\pi s/2)}\frac{1}{s^2t^2}+\frac{1}{t^2u^2}\right]\,.
\end{multline}
In the \textit{centre of momentum frame} we can choose the momenta to have the form given in \cref{eq:COM frame} 
The Mandelstam variables in the centre of momentum frame take the form,
\begin{equation}\label{eq:s-t-u-com-frame}
    s=c'E'^2\, ,\quad t=-(16+c'E'^2)\cos^2\frac{\theta}{2}\, ,\quad u=-(16+c'E'^2)\sin^2\frac{\theta}{2}
\end{equation}
Then the hard scattering limit of the four-point scattering amplitude can be expressed as,
\begin{equation}
    \mathcal{A}^{(4)}(\{p_i\})\sim (-1)^{s/2} \exp\left(-\frac{s}{2}f(\theta)\right)\, ,
\end{equation}
where,
\begin{equation}
    f(\theta)=-\sin^2\frac{\theta}{2}\ln{\sin^2\frac{\theta}{2}}-\cos^2\frac{\theta}{2}\ln{\cos^2\frac{\theta}{2}}\, .
\end{equation}

\emph{Regge limit}: in the centre of momentum frame, the Regge limit is where the scattering angle $\theta\to0$ while the variable $u$ is kept finite (see \cref{sec: scattering}). We will start with the analytically continued amplitude \eqref{eq:four-point-full-amplitude-closed}, use the mass-shell condition \eqref{eq: closed strings mass shell condition} and the Stirling's formula \eqref{eq:Stirling's Formula} to get the Regge limit as follows,
\begin{equation}
    \mathcal{A}^{(4)}(p_1,p_2,p_3,p_4)\sim\left(\frac{s}{2}\right)^{(u/2+8)}\left(1+\frac{u+16}{s}\right)^{s/2}\Gamma(-u/2)\,.
\end{equation}

\newpage
\bibliographystyle{JHEP}
\bibliography{biblio.bib}
\end{document}